\documentclass[10.5pt,a4paper]{article}
\usepackage[a4paper,top=2.6cm,bottom=2.6cm,left=2.5cm,right=2.5cm,headsep=0.8cm]{geometry}
\usepackage[T1]{fontenc}
\usepackage[utf8]{inputenc}
\usepackage{tgtermes}
\usepackage{tgheros}
\usepackage[scale=0.88]{tgcursor}
\usepackage{textcomp}
\usepackage{amssymb}
\usepackage{newunicodechar}
\newunicodechar{≤}{\ensuremath{\leq}}
\newunicodechar{≥}{\ensuremath{\geq}}
\newunicodechar{×}{\ensuremath{\times}}
\newunicodechar{−}{\ensuremath{-}}
\newunicodechar{→}{\ensuremath{\rightarrow}}
\newunicodechar{·}{\textperiodcentered}
\newunicodechar{†}{\textdagger}
\newunicodechar{²}{\textsuperscript{2}}
\newunicodechar{≈}{\ensuremath{\approx}}
\newunicodechar{±}{\ensuremath{\pm}}
\newunicodechar{α}{\ensuremath{\alpha}}
\newunicodechar{κ}{\ensuremath{\kappa}}
\usepackage{microtype}
\usepackage{setspace}
\usepackage{xcolor}
\definecolor{accent}{HTML}{1B4F72}
\definecolor{rule}{HTML}{8C9BAB}
\usepackage{graphicx}
\usepackage{booktabs,tabularx,longtable,array}
\usepackage{ltablex}\keepXColumns
\usepackage{needspace}

\usepackage[font=small,labelfont={bf,color=accent},labelsep=period,skip=6pt]{caption}
\usepackage{enumitem}\setlist{itemsep=2pt,topsep=4pt,leftmargin=1.4em}
\usepackage{titlesec}
\titleformat{\section}{\sffamily\large\bfseries\color{accent}}{}{0pt}{}
\titleformat{\subsection}{\sffamily\normalsize\bfseries\color{accent}}{}{0pt}{}
\titlespacing*{\section}{0pt}{16pt}{6pt}
\titlespacing*{\subsection}{0pt}{12pt}{4pt}
\usepackage{fancyhdr}
\fancypagestyle{plain}{\fancyhf{}\fancyfoot[C]{\sffamily\footnotesize\thepage}}
\usepackage{hyperref}
\usepackage{orcidlink}
\hypersetup{colorlinks=true,linkcolor=accent,urlcolor=accent,citecolor=accent}
\usepackage{url}
\usepackage{xurl}
\begin{document}
\thispagestyle{plain}
\begin{center}
{\sffamily\LARGE\bfseries\color{accent} Reviewing Beyond the Change: Repository Size and Existing Code Across Agent and Human Pull Requests\par}\vspace{14pt}
{\large Sabry E. Farrag\,\orcidlink{0009-0002-0735-7443}}\par\vspace{6pt}
{\small School of Architecture, Computing and Engineering, University of East London, London, United Kingdom}\par\vspace{4pt}
{\small Corresponding author: \href{mailto:sabryelgatem@gmail.com}{sabryelgatem@gmail.com}}\par
\vspace{6pt}{\small\textbf{Data and code:} \href{https://doi.org/10.5281/zenodo.23290164}{\raisebox{-0.35ex}{\includegraphics[height=1.42ex]{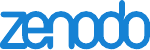}}\,Zenodo (doi:10.5281/\allowbreak{}zenodo.23290164)} \textperiodcentered{} \href{https://github.com/Gatem/beyond-the-change-replication}{\raisebox{-0.35ex}{\includegraphics[height=1.90ex]{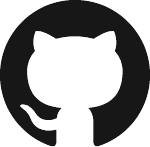}}\,GitHub} \textperiodcentered{} \href{https://huggingface.co/datasets/SabryFarrag/review-pointers-beyond-the-change}{\raisebox{-0.35ex}{\includegraphics[height=1.90ex]{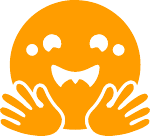}}\,Hugging Face}}\par
\end{center}\vspace{6pt}
{\color{rule}\hrule}\vspace{8pt}
\noindent{\sffamily\bfseries\color{accent} Abstract}\par\vspace{3pt}
\noindent{\small Reviewers sometimes name existing repository items outside the change under review. We ask whether such references are more common in larger repositories, and whether the pattern differs between agent-authored and human-authored pull requests. An instrument resolves names in inline review comments against the base-commit repository; size-matched negative controls estimate chance matches. Among 92,042 comments on agent pull requests in 2,034 GitHub repositories, 6.5 per cent name an existing item outside the change. An internally pre-specified log-linear model finds an odds ratio of 1.35 per tenfold increase in code files (interval 1.15 to 1.59), and size-matched negative controls indicate that chance matching does not account for the rise. Exploratory analyses suggest that the association is concentrated from small to mid-sized repositories and is imprecise among large ones. Human pull requests in 1,845 of these repositories show a similar association. A pre-specified comparison finds a small agent excess just within the bound set as negligible, but the evidence does not establish a robust agent-specific difference. The instrument misses many references, so reported shares likely understate prevalence, and the validation cannot estimate how recall changes with repository size.}\par\vspace{6pt}
\par
\vspace{8pt}\noindent{\sffamily\bfseries\color{accent} Acknowledgements}\par\vspace{3pt}
\noindent{\small We are grateful to Amelie Rosso, Aziza Tawfiq Yousif, Barry Fishel, Emilian Patrascu, Ivan Perslav, Merna Ezz Eldien, Wael Osama, Wei Fan and Youssef Saber, senior software engineers who gave this study their time and expertise. The study is better for their help.}\par
\vspace{8pt}{\color{rule}\hrule}\vspace{10pt}

\section*{1 Introduction}
\addcontentsline{toc}{section}{1 Introduction}
In Arize-ai/\allowbreak{}phoenix, an agent's pull request changed documentation on evaluating data frames, and a reviewer wrote on it: ``I think the more common pattern is to use \texttt{async\_\allowbreak{}\allowbreak{}evaluate\_\allowbreak{}\allowbreak{}dataframe} for batch evaluations so we might want to highlight that more than the base sync version.'' The function the reviewer named was already defined in a file the pull request did not touch. The reviewer was not judging the changed code alone; they were pointing the contributor to an existing repository item beyond the change.

Such comments are a visible form of project knowledge in review. A reviewer may point to an existing helper to reuse, a module whose conventions the change should follow, or a file that constrains where new code belongs. In a larger repository, the part of the codebase that a single change touches is smaller relative to the whole, so reviewers may need to make more of the surrounding code explicit. Our working hypothesis adds that a coding agent works from the part of a repository it is given or finds. It may therefore receive more such pointers than pull requests by people who work in the repository. Many agent pull requests, however, are opened under the account of a developer who often works in the repository (Section 3.2). The comparison is therefore between agent-written and other pull requests, not between agents and people working alone.

Existing work studies the expectations, outcomes and challenges of review. It shows that human review transfers knowledge of a project, and that understanding the change in its context is the reviewer's central difficulty (Bacchelli and Bird, 2013). Reviewers also share links, most often to provide context (Wang et al., 2021). Work on agent pull requests studies their acceptance, revision and reception (Li et al., 2025; Khelifi et al., 2026; Watanabe et al., 2026). It finds that agents miss chances to reuse existing code in one Python repository (Huang et al., 2026), and reads the themes of review comments on them (Haider and Zimmermann, 2026). No study directly measures whether review comments name existing repository items outside the change, relates that measure to repository size, or compares agent and human pull requests within the same repositories.

We measure all three. An instrument resolves the names in each top-level inline review comment against the repository at the pull request's base commit, the state the contributor started from. It records whether the named item lies outside the change. Size-matched negative controls estimate how often the same names match by chance. We ask four questions, in the wording of our internal plan (Section 3.11):

\begin{itemize}
\item \textbf{RQ1.} Of the inline review comments that reviewers leave on agent-authored pull requests, what share name an item of the repository that lies outside the pull request's change? A share under 1 per cent would make such references too rare to matter for review practice.
\item \textbf{RQ2.} In those comments, do the odds of naming an item outside the change rise with the size of the repository at the pull request's base commit?
\item \textbf{RQ3.} Within the same repositories, is the size gradient of outside references steeper for agent-authored than for human-authored pull requests?
\item \textbf{RQ4 (exploratory).} Is the size gradient different for agent pull requests opened when the repository carried a context file that the pull request's own agent reads?
\end{itemize}
We find that detected references to existing code outside the change are associated with repository size in both agent and human pull requests. We call a repository small below about 300 code files at the base commit, mid-sized from about 300 to 3,000 and large above about 3,000. Exploratory analyses suggest that the association is concentrated from small to mid-sized repositories. A pre-specified comparison finds a small agent excess just within the bound set as negligible, but the evidence does not establish a robust agent-specific difference. Measurement error and sensitivity to size modelling limit stronger interpretations.

Our contributions are three: an automatic measure of pointers to existing code outside the change, with negative controls for chance matches; estimates of how often such references occur and how they are associated with repository size; and a comparison of agent and human pull requests within the same repositories, with every sensitivity reported. A replication package holds the data and the scripts that recompute the estimates. Section 2 reviews related work, Section 3 describes the method, Section 4 reports the results, Section 5 discusses them, Section 6 sets out the threats to validity and Section 7 concludes.

\section*{2 Related work}
\addcontentsline{toc}{section}{2 Related work}
\subsection*{2.1 Agent-authored pull requests}
\addcontentsline{toc}{subsection}{2.1 Agent-authored pull requests}
Pull requests written by autonomous coding agents on GitHub are now numerous enough to study at scale. The AIDev dataset collects them across five agents, with their reviews and inline review comments; its later version offers a curated subset from repositories with more than a hundred stars (Li et al., 2026). Its earlier preprint compares agent pull requests with human pull requests from the same popular repositories (Li et al., 2025). Agent pull requests there are accepted less often, bot reviewers appear on them more often, and, in a single-project case study, they change code complexity less often. Pull requests created with Claude Code were merged less often than human pull requests from the same authors and projects (Watanabe et al., 2026). Those that needed revision were revised mainly for bug fixes, documentation and adherence to project-specific standards, and most rejections with a stated reason reflected project context. Developers who declare AI-generated code do so in several ways, a practice documented mostly for assistant tools that predate autonomous agents (Kashif et al., 2026). These studies describe outcomes; none measures what reviewers point to.

A second group asks why agent pull requests fail. Among rejected agent pull requests annotated by hand, most rejections come from reviewers abandoning the pull request, duplicates and failing builds or tests, and larger changes are merged less often (Ehsani et al., 2026). A taxonomy applied with a language-model classifier finds functional failures dominant among rejected pull requests with review comments, and most rejected pull requests closed without any inline review comment (Hindi et al., 2026). A dose-response analysis relates the size of the change to merge probability and review friction and finds a negative relation for every agent (Destefanis et al., 2026). Reviewer engagement is the strongest correlate of integration (Nachuma and Zibran, 2026). Another study asks which factors decide acceptance, motivated by the review effort agent pull requests place on integrators (Strauss and Schulze, 2026). Across the first four of these studies the unit is the pull request and its outcome, and size is the size of the change, not of the repository.

\subsection*{2.2 What reviewers say about agent contributions}
\addcontentsline{toc}{subsection}{2.2 What reviewers say about agent contributions}
The study closest to ours in its unit of analysis reads the review comments themselves (Haider and Zimmermann, 2026). It derives twelve themes from inline review comments on agent pull requests in AIDev's curated subset, labelled by a local language model checked against one human annotator. Functional correctness is the most frequent theme, followed by refactoring, documentation and style. No theme is dedicated to knowledge of the project beyond the change, such as existing helpers, conventions or placement. The study has no human baseline, and repository size is not a variable.

A study of developer interventions compares closed agent pull requests with human pull requests from AIDev repositories that contain both (Khelifi et al., 2026). People intervened in a smaller share of agent pull requests, and the median pull request carried no inline review comment in any group. A thematic coding places most interventions on agent pull requests at the level of guidance, including enforcing project-specific conventions and explaining project-specific context. Reviewers do, then, supply project knowledge to agents. That study codes whole interventions by theme, compares pooled shares, and does not ask whether a comment names an item of the repository outside the change or relate it to repository size.

Other studies read neighbouring signals. In one Python repository (crewAI), agents often ignore opportunities to reuse existing code, and reviewers react to their contributions with more neutral or positive emotion than to human ones (Huang et al., 2026). Developers' forum discussions of low-quality AI contributions describe a shift of work from contributor to reviewer (Baltes et al., 2026). Adoption of GitHub Copilot in open source raises contributions and also coordination time, more so for peripheral developers (Song et al., 2026). Studies of what agents change find agent refactorings that differ in kind from developers' (Ottenhof et al., 2026) and agent-written technical debt that most often records requirement debt (Brahmi et al., 2026). Code generated with ChatGPT was rarely adopted in full in pull requests (Ogenrwot and Businge, 2026). A conceptual analysis argues that agents take architectural decisions that teams seldom review as architecture; it rests on a survey of tools and a single-run case study (Konrad et al., 2026). Review comments can also be written by AI tools about human code. Such comments lead to changes less often than human ones (Sun et al., 2026), and a deployed review assistant retrieved extra code context for a language model (Olewicki et al., 2026). None of these studies measures which items of the repository reviewers point agents to.

\subsection*{2.3 What review needs from outside the change}
\addcontentsline{toc}{subsection}{2.3 What review needs from outside the change}
Modern code review is an established research area (Davila and Nunes, 2021). A recurring finding is that review is mostly not about functional defects. Most defects found in industrial and student reviews make code easier to understand and change rather than affect visible functionality (Mäntylä and Lassenius, 2009). In two open-source projects, evolvability changes outnumber functional ones among review-driven changes, and in one of them tasks that alter more files and more code receive more review-driven changes (Beller et al., 2014). At Microsoft, reviews deliver knowledge transfer, team awareness and alternative solutions, and understanding the change is the reviewer's central difficulty (Bacchelli and Bird, 2013). These studies establish why review needs knowledge of the project, but not how that need varies with the project's size.

That understanding depends on knowledge a change does not carry. Reviewers' questions in three Gerrit projects show information needs that include the context into which a change is plugged (Pascarella et al., 2018). Reviewers also share links to meet such needs (Wang et al., 2021), and reviews link to other reviews (Hirao et al., 2019). Many coding conventions are unwritten and emerge within a codebase, so inferring them requires a global view of it (Allamanis et al., 2014). These studies analyse questions and links as written. They do not resolve the names a comment writes as code, ask whether the named item lies outside the change, or relate either to repository size.

Who reviews matters for what review delivers. The share of useful comments rises over a reviewer's first year and falls as the change touches more files (Bosu et al., 2015). Many developers review modules they never authored, and the share of reviewers without expertise in a module is strongly related to its post-release defects (Thongtanunam et al., 2016). Reviewer workload, experience and participation are associated with review quality (Kononenko et al., 2015), and low review coverage with more post-release defects (McIntosh et al., 2014). These studies concern human changes only, and none relates the content of comments to repository size.

\subsection*{2.4 Architecture, repository size and the context available to agents}
\addcontentsline{toc}{subsection}{2.4 Architecture, repository size and the context available to agents}
Some of the knowledge a reviewer supplies concerns the system as a whole. Review comments in OpenStack raise symptoms of architecture erosion, such as duplicated functionality, and most are addressed (Li et al., 2022). Developers discuss the architectural impact of changes that significantly alter the module structure in only about a third of cases (Paixao et al., 2021). Practitioners often check architecture consistency informally (Ali et al., 2018), although static conformance techniques can encode such rules (Passos et al., 2010). Some architectural debt spreads to other parts of the system (Martini and Bosch, 2015), so a locally correct change can still cost the system. In Java repositories that adopted agents, density-normalised architectural outcomes mislead when the treatment changes system size (Larsen and Moghaddam, 2026). None of these studies relates review comments on agent pull requests to repository size.

Two studies place agent pull requests in their repositories. Multilevel models of integration friction compare agent with human pull requests in the same repositories, with codebase size and age as covariates (Russo, 2026). The repository share of variance stays large after these controls and is larger for agent pull requests. That design is the closest precedent for our human baseline, but it measures process friction, not what reviewers write. Agent pull requests are also merged less often in more popular repositories (Bhavnagri, 2026); that study measures popularity by stars, studies outcomes and has no human baseline. Stars and code size can vary together, so a size gradient across repositories risks confounding by repository. Agent use itself varies with size: traces of coding agents are found more often in larger projects than in the smallest ones (Robbes et al., 2026). Repositories of different sizes therefore receive different amounts of agent work, a reason to hold the repository fixed when comparing agent and human pull requests. Comments are also clustered within repositories; cluster-robust tests can over-reject with few clusters (Cameron et al., 2008), and our analyses have far more.

Arguments about where agents struggle point to the state of the codebase. A multivocal review of sources of mixed standing argues that as projects grow an agent loses awareness of distant modules (Farrag, 2026), the size mechanism tested here, but gives no gradient from data. Resolving real GitHub issues often requires coordinated edits across files (Jimenez et al., 2024). Systems built for repository-level tasks make finding the relevant code an explicit step, by search and navigation tools (Yang et al., 2024), localisation before repair (Xia et al., 2025) or planning across dependent files (Bairi et al., 2024). Retrieving repository context improves code completion over an in-file baseline (Zhang et al., 2023). Project knowledge can also be written down for agents in context files, which most often cover conventions and architecture (Mohsenimofidi et al., 2026; Chatlatanagulchai et al., 2026). Such files change benchmark task success little while raising cost (Gloaguen et al., 2026), and adding one raises the merge rate in some projects and lowers it in others (Arabat and Sayagh, 2026). These are benchmark, outcome or content studies; none reads review comments or varies repository size. Our exploratory RQ4 asks whether the size gradient differs when the agent's own context file is present.

\subsection*{2.5 Language models as annotators}
\addcontentsline{toc}{subsection}{2.5 Language models as annotators}
Labels from language models stand in for human coding in several of the studies above (Haider and Zimmermann, 2026; Hindi et al., 2026; Sun et al., 2026). Agreement between models and people varies strongly by software engineering task, and low agreement between models flags the tasks where models also disagree with people (Ahmed et al., 2025). Model judges generalise poorly across code tasks (Wang et al., 2025). The alternative annotator test states when a model may replace human annotators, and it requires at least three human annotators (Calderon et al., 2025). Krippendorff's alpha has been proposed as the standard measure of agreement among several coders (Hayes and Krippendorff, 2007). Model labels for a new construct therefore need their own validation, so we treat labels from model judges as a sensitivity until people check them.

\subsection*{2.6 Linking text to code elements}
\addcontentsline{toc}{subsection}{2.6 Linking text to code elements}
Deciding whether a comment names an item of the repository is a traceability problem with a long history. Light lexical rules link e-mails to classes and functions with a precision that differs between systems, and class names that are single dictionary words are the most problematic (Bacchelli et al., 2010). In documentation and support messages, most false links come from terms that share a name with a concept, and context-based filters largely resolve the ambiguity (Dagenais and Robillard, 2012). An extractor for forum posts requires an unqualified single-word method name to carry a parenthesis or a type qualifier, because pervasive names are hard to link (Rigby and Robillard, 2013). Each of these studies reports precision per system, and none relates the rate of chance matches to the size of the codebase. A larger tree holds more names, so a measure that resolves names against the repository can match more often in a larger repository by chance alone. A size gradient in such a measure must therefore be read net of chance matching estimated at each size.

\subsection*{2.7 Position of this study}
\addcontentsline{toc}{subsection}{2.7 Position of this study}
Studies of agent pull requests measure acceptance, failure, change size, review friction, developer interventions, reviewer sentiment and the themes of review comments. Studies of human review show that reviewers bring knowledge of the wider system to a change. Work on context files shows that developers try to write that knowledge down for agents, and traceability research shows how links from text to code go wrong. No study measures, in review comments on agent pull requests, whether reviewers name repository items that exist at the base commit and lie outside the change. None relates that measure to repository size, or compares it with human pull requests in the same repositories. We do all three. The method must therefore resolve each comment against its repository at the base commit, estimate chance matches at each size, and compare agent and human pull requests within the same repositories.

\section*{3 Method}
\addcontentsline{toc}{section}{3 Method}
This section sets out which review comments we study, how we decide whether a comment names an item outside the change, and how we answer each question. RQ1 to RQ3 are confirmatory and RQ4 is exploratory. A deterministic instrument, a fixed set of rules applied by code, measures every eligible comment. No comment is labelled by a person or a model in the main analysis. Every step, from listing pull requests to the decision tests, runs as a configured analysis step. The replication package holds the data, the arm rules, the seeds, and scripts with their configuration that recompute every estimate, and the numbers that rest on them, from the analysis table.

\subsection*{3.1 Data source and frame}
\addcontentsline{toc}{subsection}{3.1 Data source and frame}
We start from a fixed list of repositories, collect their pull requests over a fixed window, and count every comment we exclude on the way to the analysis. The list, which we call the frame, holds 4,138 GitHub repositories. We built it from the agent arm of the AIDev dataset (Li et al., 2025, 2026) in the earlier round of our study. That round was an earlier collection and analysis of the same study, whose estimates were seen before the analysis plan was frozen (Section 3.11). We read its curated subset as downloaded on 2026-10-05 and its full tables of release v4 as downloaded on 2026-10-07.

The frame holds the 1,690 repositories of AIDev's curated subset that had agent review comments; that subset kept repositories with more than 100 stars. On 2026-10-08 we added 2,451 other AIDev repositories with at least 100 stars. Of these, 1,420 had at least three agent pull requests since December 2024. The other 1,031 had one or two agent pull requests whose records had already been collected when we stopped adding such repositories. Collection ran in alphabetical order, so these are the alphabetically first of them. Three of the added repositories are not in the final list, which holds 4,138. The enlargement gave RQ2 more repositories (D1, Appendix A). Of these repositories, 4,088 were available and 50 were not.

For each available repository, we listed through the GitHub API every pull request created from 2024-12-01 up to, but not including, 2026-10-01. We also collected the review threads of the selected pull requests. We imported the collection of the earlier round into the format of the present analysis, re-applied the selection rule to it and compared the re-applied selection with the recorded one.

Two steps check that no repository's listing is incomplete. First, GitHub's pull request listing stopped at 20,000 entries for 31 repositories. For these, we completed the older part of the window by date-sliced search. We then re-drew their selection over all their pull requests with the same rule and seed, so no repository enters the analysis with a capped listing. Second, we checked completeness per repository against GitHub's search count of pull requests created in the window. Over the 4,088 repositories checked, the shortfall was 0.30 per cent, against a threshold of 1 per cent.

We count every exclusion, in this order. Pull requests in neither arm are excluded (Section 3.2), and pull requests that are not selected contribute no comments. A comment is not eligible when it is not a top-level inline review comment, when a bot account wrote it, or when the pull request's own author wrote it (Section 3.3). Comments addressed to an agent stay in the data but leave the primary analyses.

Of the 486,378 eligible comments that entered the analysis frame, we dropped 5 duplicate records of the same comment. We dropped 60 comments the instrument could not measure (a status other than ok, Section 3.4). We also dropped 2,424 comments whose repository had zero or unknown size at the base commit, 197 in the agent arm and 2,227 in the human arm. The size predictor is the logarithm of a file count, so these comments cannot enter any model with size: their trees hold no file with a code extension, or their size is unknown. The instrument found an outside reference in 1.0 per cent of the dropped agent-arm comments and 0.3 per cent of the dropped human-arm comments.

The remaining 483,889 comments form the analysis frame. Each question uses part of it:

\begin{itemize}
\item RQ1 and RQ2: 92,042 agent-arm comments not addressed to an agent, in 2,034 repositories;
\item RQ3: a comparison set of 348,721 comments in 1,845 repositories;
\item RQ4: 85,183 comments.
\end{itemize}
\subsection*{3.2 Arms and selection}
\addcontentsline{toc}{subsection}{3.2 Arms and selection}
We sort pull requests into an agent arm and a human arm by the marks that agents leave. We call the second arm human for brevity; it is strictly the arm without detected agent markers. We then select the pull requests whose comments we study.

A pull request belongs to the agent arm when AIDev labels it with an agent. Where AIDev has no label, it belongs to the agent arm when any of the following holds:

\begin{itemize}
\item its author is one of Copilot, devin-ai-integration[bot], google-labs-jules[bot], cursor[bot] or claude[bot];
\item its head branch starts with copilot/\allowbreak{}, devin/\allowbreak{}, cursor/\allowbreak{}, codex/\allowbreak{} or claude/\allowbreak{};
\item its description carries a Codex task link or the signature ``Generated with Claude Code'' (matched case-sensitively).
\end{itemize}
Like AIDev (Li et al., 2025), we identify agents by bot accounts and branch prefixes. The agent arm has six groups: Copilot, Devin, Google Jules, Cursor, OpenAI Codex and Claude Code.

Agent pull requests reach GitHub in two ways. Copilot, Devin and Google Jules open them from the agent's bot account. Claude Code and OpenAI Codex pull requests, and most Cursor ones, are opened under the developer's own account. These carry 48,792 of the RQ1 and RQ2 comments (53 per cent). Their author is then often an owner, member or collaborator of the repository. So it is for 43 per cent of the comments on Claude Code, 38 per cent on OpenAI Codex and 51 per cent on Cursor pull requests. The shares are 1 per cent on Copilot and 7 per cent on Devin pull requests. The agent arm therefore holds agent work submitted by a bot account and agent work submitted by a person who often works in the repository.

A pull request belongs to the human arm when its author is a user account whose login does not end in [bot], and its description, head branch and title carry no agent or AI-tool marker. The marker is a regular expression given with the arm rules in the replication package. Pull requests in neither arm are excluded.

We select every agent pull request with at least one review thread. In the human arm, we draw a random sample of up to 100 pull requests with at least one review thread per repository (seed 20261007). For each selected pull request we recorded its additions, deletions and changed files.

\subsection*{3.3 Eligible comments}
\addcontentsline{toc}{subsection}{3.3 Eligible comments}
We measure the comment that opens a review thread on a line of code, written by someone other than the pull request's author. This unit of analysis, a top-level inline review comment, is the first comment of a review thread on a selected pull request, created before the end of the window. A comment is eligible when its author is a user account whose login does not end in [bot], the author is not the pull request's author, and the pull request belongs to an arm.

Some comments speak to the agent instead of reviewing its change. A comment whose text matches @(copilot|codex|claude|devin|cursor|jules), ignoring case, is marked as addressed to an agent. Such a comment is an instruction to the agent's tool, part of driving the agent, rather than review of its change. We keep addressed comments in the data, exclude them from the primary analyses and restore them in sensitivity analyses.

On a pull request opened by an agent's bot account, the developer associated with the request is not its author; we call this developer the delegator. We identify delegators from the GitHub API by a broad rule: the pull request's assignees, the users whose review was requested or who requested it, and the account in a ``Requested by'' line of its description. This broad rule also catches reviewers who were only asked to review: 29 per cent of the comments it marks are not marked by the narrow rule, which keeps assignees and the ``Requested by'' account only. The main analyses keep the delegators' comments. The sensitivities that leave them out use the broad rule, which includes requested reviewers; the narrow rule is a further sensitivity.

The RQ1, RQ2 and RQ4 population is the eligible agent-arm comments not addressed to an agent. The RQ3 comparison set adds the eligible human-arm comments of the same repositories that are not addressed to an agent. It keeps only repositories with comments in both arms. It caps the human comments at 300 per repository by a seeded random flag (seed 20261007), so that repositories with heavy human review do not dominate the comparison.

\subsection*{3.4 Measuring outside references}
\addcontentsline{toc}{subsection}{3.4 Measuring outside references}
\textbf{In plain terms.} The outcome of every question is whether a review comment names an item of the repository that lies outside the pull request's change. The instrument picks out the words in a comment that are written as names. A name can be a link to a file of the repository, a file or directory path, a module name, or the name of a function, class or other definition written as code. It looks each name up in the repository as it stood at the pull request's base commit, the state the contributor started from. A comment counts as naming an item outside the change when at least one of its names resolves, under rules that reject names matching too many items, to an item that is not among the files the pull request had changed by the time the comment was written. The paragraphs below give the rules in full.

\textbf{Repository index.} To know what exists and what the pull request changed, we build an index for every pair of a pull request's base commit and the commit a comment was written on. The index holds the full tree at the base commit, the change, and the definitions of symbols. Paths are resolved against the base-commit tree; symbol candidates are retrieved from a head-derived index restricted to paths present at the base commit.

\begin{itemize}
\item The change is the set of files that differ between the base commit and the comment's commit (git diff with rename detection off) and that the pull request itself changes, as listed by GitHub's files of the pull request. The second condition matters for a branch that merged its base branch. Such a branch also differs from the base commit in every file changed upstream since, and those files are not the contributor's change. Where GitHub does not return a pull request's full file list, the change is the difference alone.
\item Definitions (def, class, function, struct, enum, type, const, export and similar declarations) are read from the code files of the default branch's head, files up to 512 kB, and restricted to files that exist at the base commit.
\item When the comment's commit cannot be fetched, we use the pull request's head and record the fallback; at most 5 per cent of pairs may fall back.
\end{itemize}
In the validation sample, the matched symbol occurred in the matched file at the base commit for 214 of 228 symbol matches (93.9 per cent); a flagged comment whose matched items all fail the factual checks, this one included, counts as incorrect (Section 3.8). This check supports the symbol index but does not establish that every symbol match in the full frame reflects a symbol present at the base commit.

\textbf{Candidates.} The instrument first collects candidate names from the comment text outside fenced code blocks. These are GitHub blob and tree URLs of the same repository, path-like strings and file names with a listed extension, dotted module names, and symbols. A symbol (a CamelCase or snake\_\allowbreak{}case name, a called name, or a single identifier) is a candidate only when written as code in backticks or written as a call. Stop words are dropped. Candidates that appear in the commented diff hunk, or equal the commented file, are skipped.

\textbf{Resolution.} Resolution decides whether a candidate names a real item of the repository and whether that item lies outside the change. A candidate counts as an outside reference when it resolves to an item of the tree at the base commit, outside vendored directories, and that item lies outside the change. It resolves when it is one of the following:

\begin{itemize}
\item an exact path or directory;
\item a bare file name carried by at most 3 paths;
\item a unique path suffix;
\item a module path;
\item a symbol of at least 4 characters that is not on a list of generic words and is defined in at most 5 code files present at the base commit.
\end{itemize}
The item lies outside the change when it is not among the files changed between the base commit and the comment's commit. A directory counts as changed when any changed file lies in it. We set these limits by the lesson of earlier work on linking developer text to code. Names that are also common words produce false links (Bacchelli et al., 2010; Dagenais and Robillard, 2012), and compound names are easier to link correctly than pervasive single words (Rigby and Robillard, 2013). Rigby and Robillard (2013) accept an unqualified single-word method name only with a parenthesis or a type qualifier. We apply the same idea by accepting a single identifier only as code or as a call.

\textbf{Outcome.} The outcome ref\_\allowbreak{}out is 1 when at least one candidate of the comment counts and 0 otherwise. Each hit records its kind (URL, path, symbol) and whether the item is a process file (CHANGELOG, CONTRIBUTING, CODEOWNERS, workflow files, Makefile and similar). Each comment receives a status: ok, no index, no base commit, or change unknown. We analyse only comments with status ok, and at least 95 per cent of eligible comments must reach it. We set the rule thresholds on a development set of the earlier round and fixed the instrument code by hash at the pilot gate (Section 3.10).

\textbf{Repository size and covariates.} Besides the outcome, the models use the following variables.

\begin{itemize}
\item Repository size is the number of files with a code extension in the full tree at the pull request's base commit, read from the same index; the count includes vendored directories. The size predictor lsize is its base-10 logarithm, so one unit is a tenfold increase. Every model centres it on its mean over the analysis frame, which corresponds to about 845 code files.
\item A reviewer is an insider when the comment author's association with the repository is owner, member or collaborator, as GitHub reported it at collection (Appendix F.1).
\item PR size, the size of the pull request (PR), is log10(additions + deletions + 1), and month is the month of the pull request's creation.
\item For RQ4, context\_\allowbreak{}file is 1 when the context file read by the pull request's own agent exists in the tree at the base commit. A context file is a file of instructions in the repository that the agent reads, as the agent's public documentation names it: \texttt{CLAUDE.\allowbreak{}md} anywhere for Claude Code; \texttt{.\allowbreak{}github/\allowbreak{}\allowbreak{}copilot-\allowbreak{}instructions.\allowbreak{}md} or \texttt{.\allowbreak{}github/\allowbreak{}\allowbreak{}instructions/\allowbreak{}\allowbreak{}*.\allowbreak{}instructions.\allowbreak{}md} for Copilot; \texttt{AGENTS.\allowbreak{}md} anywhere for OpenAI Codex and Google Jules; \texttt{.\allowbreak{}cursorrules} or \texttt{.\allowbreak{}cursor/\allowbreak{}\allowbreak{}rules/\allowbreak{}\allowbreak{}} for Cursor. No context file is attributable to Devin alone, so we leave Devin pull requests out of RQ4. We adapt the agent-specific file names of Arabat and Sayagh (2026). They count AGENTS.md as common to several agents and list a pull-request template for Devin; we assign AGENTS.md to the agents that read it and use no file for Devin. Chatlatanagulchai et al. (2026) describe such context files for Claude Code, Codex and Copilot.
\end{itemize}
\subsection*{3.5 Negative controls and net-of-chance contrasts}
\addcontentsline{toc}{subsection}{3.5 Negative controls and net-of-chance contrasts}
A name can resolve in a large repository by chance, and larger trees offer more names to match. We estimate such chance matching with two negative controls (Lipsitch et al., 2010), under the assumption stated in Appendix F.4. Each control resolves a comment's candidates, with the same rules, against a different repository of similar size, where a match cannot be a reference to the reviewed project.

\begin{itemize}
\item \textbf{Next partner (as pre-specified).} The partner is the next repository in size order among frame repositories of another owner. The partner is read at its indexed default-branch head, and nothing is excluded as changed. The outcome is placebo\_\allowbreak{}ref.
\item \textbf{Random band (robustness).} For each repository, we draw five partners at random (seed 20261009, fixed before the run) among frame repositories of another owner with the same primary language, within a band of size around its own. The band is the narrowest of 15, 30, 50 and 100 per cent of size that offers five such partners. Where no width offers five, the widest width that offers any partner is used. The width used is recorded with each comment (band\_\allowbreak{}matched). Partners are read at their heads. The outcome placebo\_\allowbreak{}band is the mean match over the partners drawn.
\end{itemize}
Both controls pair repositories by the number of paths at the indexed head. At most 5 per cent of repositories may lack a partner, and at most 20 per cent may have fewer than five band partners. The controls estimate chance matching under their pairing assumptions; they do not correct for all forms of dependence between repositories, such as shared ancestry or common libraries.

``Net of chance'' means the measured gradient minus the gradient of the chance matches. For each contrast we fit a linear probability model, a linear regression of a yes-or-no outcome, on ref\_\allowbreak{}out and on the placebo outcome. We fit them separately and once more as a single fully interacted stacked fit. The net coefficient is the measured coefficient minus the chance coefficient, in percentage points per tenfold increase in size. Our code checks on unrounded values that the stacked fit reproduces this difference.

The RQ2 contrast (rq2\_\allowbreak{}net) regresses each outcome on lsize alone in the agent arm. The RQ3 contrast (rq3\_\allowbreak{}net) regresses each outcome on arm, lsize, their interaction and reviewer insider status. It includes repository fixed effects, a separate baseline for each repository, estimated separately for each outcome. It reports the interaction, which measures how much the size gradient differs between the arms. These two contrasts with the next-partner control are the net-of-chance companions of RQ2 and RQ3. They are linear models of their own, so their percentage-point estimates are not conversions of the odds ratios of Section 3.6. The same contrasts with the random-band control, and restricted to repositories matched within 15 per cent of size, are sensitivities (Appendix B).

\subsection*{3.6 Pre-specified models and decision rules}
\addcontentsline{toc}{subsection}{3.6 Pre-specified models and decision rules}
Three logistic models, fixed in the plan, answer RQ1 to RQ3, and only they enter the decision tests. ``Pre-specified'' means fixed before the present full-frame run, not before any related outcome data had been examined. A logistic regression models the odds that a comment names an outside item. For size, its result is an odds ratio per tenfold increase in code files; above 1, such comments are more likely in larger repositories. We fit all models on the analysis frame with standard errors clustered by repository, which allow comments of the same repository to be correlated.

\begin{itemize}
\item \textbf{RQ1 (rq1).} An intercept-only logistic regression of ref\_\allowbreak{}out on the agent-arm population. Its intercept, transformed to a share, estimates the share of comments that name an item outside the change.
\item \textbf{RQ2 (rq2\_\allowbreak{}registered).} A logistic regression of ref\_\allowbreak{}out on lsize and reviewer insider status, with fixed effects for the agent group and the month, on the agent-arm population. The odds ratio of lsize is the change in the odds of an outside reference per tenfold increase in size.
\item \textbf{RQ3 (rq3\_\allowbreak{}registered).} A logistic regression of ref\_\allowbreak{}out on the arm indicator, lsize, their interaction and reviewer insider status, on the RQ3 comparison set. The exponentiated interaction is the ratio of the agent to the human size odds ratio; above 1, the agents' rise is steeper. Repository effects enter through the Mundlak device (Mundlak, 1978; Chamberlain, 1980; Wooldridge, 2019). The device adds each repository's averages of the arm indicator, lsize, their interaction and reviewer insider status as extra terms. The arm and size terms then describe differences within repositories. In a logistic model this is an approximation that also draws on variation between repositories. It is not a random-effects model, and it does not remove every repository characteristic.
\end{itemize}
The plan fixed a smallest effect of interest (SESOI) for each test, below which an effect is too small to matter for review practice. It is an odds ratio of 1.15 per tenfold increase for RQ2, and a ratio of odds ratios of 1.25 (or 1/\allowbreak{}1.25) for RQ3. On the percentage-point scale, both correspond to about one outside reference per hundred comments per tenfold increase. Appendix D gives the source of each bound. The confirmatory models keep the identifiers rq2\_\allowbreak{}registered and rq3\_\allowbreak{}registered from the frozen plan's files; the plan was not publicly registered.

The plan states the form of each answer:

\begin{itemize}
\item RQ1: a share with an interval; the plan sets 1 per cent as the share below which outside references would be too rare to matter for review practice;
\item RQ2: an odds ratio per tenfold size, above 1 in support of the bounded-view account;
\item RQ3: a ratio of odds ratios, which the plan reads, within the bound, as the gradient belonging to large codebases for everyone, a reading we qualify in Section 5.2;
\item RQ4: an exploratory ratio.
\end{itemize}
We test each confirmatory term two ways: whether it differs from zero, and whether it is smaller than the SESOI. The second is an equivalence test. Two one-sided tests (TOST; Lakens, 2017) on the odds-ratio scale ask whether the effect lies within the bounds 1/\allowbreak{}SESOI and SESOI, at alpha 0.05, and are reported with the 90 per cent interval. The two-sided p values of RQ2 and RQ3 are adjusted by the Holm procedure (Aickin and Gensler, 1996), which corrects for running more than one test. Both p values are kept separately in the output. RQ1 has a single estimate and no test.

The verdict combines the two tests into one of four outcomes:

\begin{itemize}
\item ``different from zero'': Holm-adjusted p < 0.05 only;
\item ``equivalent to zero within the smallest effect of interest'': TOST p < 0.05 only;
\item ``different from zero, and smaller than the smallest effect of interest'': both;
\item ``inconclusive'': neither.
\end{itemize}
These labels are the pre-specified wording; zero refers to the log-odds scale. On the odds-ratio scale, zero corresponds to an odds ratio of 1, and outside the verdict labels we state whether an interval includes 1.

Table 1 gives the role of each kind of estimate; Table B1 in Appendix B maps every estimate to its role. Only the three pre-specified models carry a decision.

\needspace{20\baselineskip}
{\small
\setlength{\tabcolsep}{4pt}
\begin{tabularx}{\linewidth}{XX}
\caption{The role of each kind of estimate.}\\
\toprule
\textbf{Kind} & \textbf{Role} \\
\midrule\endfirsthead
\toprule \textbf{Kind} & \textbf{Role} \\ \midrule\endhead
Pre-specified models (rq1, rq2\_\allowbreak{}registered, rq3\_\allowbreak{}registered) & decision tests for RQ2 and RQ3; RQ1 a share \\
Net-of-chance companions (next-partner control) & the same questions on the linear probability scale \\
Chance and robustness controls (random band, re-paired partner) & sensitivities of the net contrasts \\
Error correction (DSL) & sensitivity to the instrument's error \\
Validity checks (known groups, convergent check) & what the measure tracks \\
Sensitivities planned (D11) or added after the results (D19, D20, D22) & how the answers depend on the specification; RQ4 exploratory \\
\bottomrule
\end{tabularx}}
As a check on the computation, we fit every model with two implementations, the Python packages pyfixest (version 0.60.0) and statsmodels (Seabold and Perktold, 2010), and check the nesting of models against the data. The two pre-specified models with covariates require both implementations on the full frame, and both agree on them, as they did in the pilot that passed the gate. We fit the models with repository fixed effects with pyfixest alone on the full frame; both implementations agreed on them in the pilot.

\subsection*{3.7 Sensitivity and exploratory analyses}
\addcontentsline{toc}{subsection}{3.7 Sensitivity and exploratory analyses}
None of the analyses beyond the pre-specified models enters the decision tests. Some were planned (D11); most were added after the full-frame estimates were seen, in response to language-model reviews (D19, D20, D22). They show how the answers depend on the specification, not a preferred result. We apply the pre-specified decision rule to each RQ3 sensitivity for description only. The main text reports the following.

\begin{itemize}
\item \textbf{RQ2.} Beside the pre-specified model, its net-of-chance companion (Section 3.5) and the functional form of the gradient: a term for the square of lsize, the model fitted below and at or above 1,500 code files, and the model on repositories with at least 300 code files.
\item \textbf{RQ3.} Beside the pre-specified model, the repository-month comparison. It compares agent and human pull requests reviewed in the same repository in the same month. It has one fixed effect per repository-month cell, keeps size as a term, and holds pull-request size and the agent group fixed.
\end{itemize}
Appendix B describes every other analysis, and Table B2 reports the RQ3 specifications; Appendix F.4 notes the pre-specified model on the fixed-effects sample. These vary the terms of the model, the period, pull-request size, the reviewers and authors included, the unit of analysis, the repository's traits and the curve fitted to size. RQ4 is a logistic regression of ref\_\allowbreak{}out on lsize, context\_\allowbreak{}file, their interaction and reviewer insider status, on the agent-arm population without Devin. It carries no decision rule.

\subsection*{3.8 Validation of the instrument}
\addcontentsline{toc}{subsection}{3.8 Validation of the instrument}
We assessed how often the instrument correctly identifies references to items outside the change, and how often it misses them. Precision is the share of the instrument's hits that are true references. Recall is the share of true references that it finds. The false-omission rate is the share of unflagged comments that do name an outside item. The sample holds 600 comments, stratified by arm, repository-size third and the instrument's verdict. Every estimate is weighted by the inverse of each comment's inclusion probability.

For each flagged comment, up to three matched items were checked against the GitHub API, a source the instrument does not read. The checks were that the item existed at the base commit, that a matched symbol occurred in the file it resolved to, and that the item lay outside the change. Three language models from different families, blind to arm and size, judged whether the comment refers to the matched item and whether it directs the contributor to it. For an unflagged comment, they judged whether it names an item outside the commented hunk. The author and four software engineers named in the acknowledgements checked part of the model labels, including labels reversed on purpose. Where people answered, their label took precedence.

A flagged comment is correct when one matched item passes the factual checks and the final judgement confirms the reference. An unflagged comment is a miss when the final judgement names an item outside the commented hunk that resolves in the index to an item outside the change. We report recall also under a second rule that counts every miss the judgement reports, confirmed by the index or not. We report weighted precision, recall and false-omission rate overall, by arm and by repository-size third.

The validation establishes whether the instrument identifies references correctly. It does not establish whether a reference reflects a need for project knowledge. Where the reference rests on model judgements, the measures are agreement with the models. The validation supports the instrument's flagged references but cannot establish how recall varies with repository size. Appendix C gives the sample's quotas, the judges, the scoring rules and the reading of the sheets in full.

Two further checks ask what the measure tracks. The known-groups check (Cronbach and Meehl, 1955) asks whether the measure separates reviewers who should differ. We regress ref\_\allowbreak{}out on reviewer insider status in each arm (kg\_\allowbreak{}agent, kg\_\allowbreak{}human). Insiders are expected to name items outside the change more often, an odds ratio above 1, and we report an odds ratio at or below 1 as a failed check. These two models carry no repository effects, so they compare reviewers across repositories of different sizes. As a sensitivity reported beside them, which does not change their verdict, we fit both with repository fixed effects (kg\_\allowbreak{}agent\_\allowbreak{}fe, kg\_\allowbreak{}human\_\allowbreak{}fe).

The convergent check (D16) tests convergent validity in the broad sense of agreement with an independent indicator (Campbell and Fiske, 1959). It asks whether the files a flagged comment names, outside the change when it was written, are among the files the pull request changes by its end. The baselines are as many files drawn at random from the base commit's tree and, for each named file, a file drawn from the same directory (seed 20261010). The second baseline matters because files next to a change are changed together more often whatever a reviewer writes. Intervals come from a cluster bootstrap over repositories.

\subsection*{3.9 Error correction as a sensitivity}
\addcontentsline{toc}{subsection}{3.9 Error correction as a sensitivity}
If the instrument errs more often in one arm than the other, the error can create or hide a difference in gradients. We therefore correct the estimates of RQ2 and RQ3 for the instrument's measured error by design-based supervised learning (DSL; Egami et al., 2023). DSL uses the validation sample as the reference, together with its known inclusion probabilities.

The DSL model for RQ2 regresses ref\_\allowbreak{}out on lsize and reviewer insider status in the agent-arm population. The DSL model for RQ3 regresses it on the arm indicator, lsize, their interaction and reviewer insider status in the comparison set, with repository means of the arm indicator and lsize in place of fixed effects. Neither is the pre-specified specification. The RQ2 model has no agent or month effects, and the RQ3 model carries only the repository means of the arm and size. Our DSL implementation, in the replication package, fits a pooled model without fixed effects. Agent and month effects, though identified from all rows, would leave the correction, which rests on a few hundred labelled comments, too imprecise to read. A second RQ3 correction carries all four Mundlak means of the pre-specified model. The reference rests on people's labels where they answered and on model judgements elsewhere. We therefore report the DSL estimates as a sensitivity beside the measured estimates, and they do not replace them.

\subsection*{3.10 Pilot and gate}
\addcontentsline{toc}{subsection}{3.10 Pilot and gate}
Before any full-frame collection or measurement, we ran the whole analysis on a pilot of 60 seeded repositories, at most 5 per cent of the frame. The full run could proceed only after the pilot passed a set of checks, the gate. The gate required, among other things, an independent validation, two implementations of every model, completeness, and a review by model families other than Claude (Appendix A lists every check). Its last check is a lock that records the hashes of the instrument, the estimators and the configurations. The pilot passed the gate on 2026-10-09, before any full-frame outcome had been computed. Later stages refuse to run while any locked file differs from its pilot hash, so the full-frame analysis runs only on files that have passed the gate. The pilot's estimates are checks of the instrument and the analysis steps, not results of the research questions.

\subsection*{3.11 Deviations from the pre-specified plan}
\addcontentsline{toc}{subsection}{3.11 Deviations from the pre-specified plan}
The analysis plan was frozen internally on 7 and 8 October 2026, before the present full-frame run, after estimates from an earlier round of the study had been seen; it was not publicly registered. We report every departure, including the analyses added after the full-frame estimates were seen (D19, D20 and D22), in Appendix A.

\section*{4 Results}
\addcontentsline{toc}{section}{4 Results}
This section answers the research questions in turn and then reports how well the instrument measures. RQ1 and RQ2 rest on 92,042 eligible comments on agent pull requests in 2,034 repositories. RQ3 compares agent and human comments in the 1,845 repositories that have comments in both arms, 348,721 comments in all. Unless stated otherwise, every interval is a 95 per cent interval with standard errors clustered by repository.

\subsection*{4.1 How often reviewers point outside the change (RQ1)}
\addcontentsline{toc}{subsection}{4.1 How often reviewers point outside the change (RQ1)}
About one comment in fifteen on agent pull requests names existing code outside the change. Such comments make up 6.5 per cent of the eligible comments on agent pull requests (interval 5.6 to 7.5). With the comments addressed to an agent restored, the share is 6.0 per cent. Weighting each repository equally rather than each comment, it is 4.7 per cent. Appendix F.1 gives the addressed comments by agent.

A comment from cybersemics/\allowbreak{}em shows what is counted. An agent's pull request changed src/\allowbreak{}actions/\allowbreak{}swapParent.ts, and the reviewer's entire comment was ``Use \texttt{rootedParentOf} util function.'' Our instrument resolved the name to src/\allowbreak{}selectors/\allowbreak{}rootedParentOf.ts. That file existed at the base commit, and the pull request did not touch it.

Most of these comments name code. Of the comments with an outside reference, in both arms, 58.2 per cent name a function, class or other defined symbol. A file or directory path is named in 33.4 per cent, and a link to a file of the repository in 16.1 per cent. A comment can name items of several kinds, so these shares sum to more than 100. Process files, such as CHANGELOG or workflow files, account for 2.1 per cent.

The share is far above chance. Looked up in unrelated repositories of similar size, where any match is a coincidence, the same names match 1.1 per cent of the comments with the next partner and 1.4 per cent with the random band. The measured share is above the 1 per cent that the plan set as too rare to matter, and so is the share net of chance.

\subsection*{4.2 Repository size and outside references in the agent arm (RQ2)}
\addcontentsline{toc}{subsection}{4.2 Repository size and outside references in the agent arm (RQ2)}
A pre-specified log-linear model finds that detected outside-change references are more likely in larger agent-arm repositories. The model adjusts for reviewer insider status and holds the agent group and the month fixed. In it, the odds that a comment names an outside item rise by an odds ratio of 1.35 per tenfold increase in code files at the base commit (interval 1.15 to 1.59).

The pre-specified decision rule finds the rise different from zero. The Holm-adjusted p value is below 0.001. The equivalence test against the bound of 1.15 gives p = 0.972 (90 per cent interval 1.18 to 1.55), so it does not find the rise negligible. The pre-specified verdict is therefore ``different from zero''.

The rise remains net of chance matches, as the size-matched controls estimate them. On the percentage-point scale, the measured share rises by 2.46 points per tenfold increase in size, and the share matched by chance by 0.57. Net of chance, the rise is 1.89 points per tenfold increase (interval 1.06 to 2.72). With the adjustments of the RQ2 model, the net rise is 1.11 points (interval 0.43 to 1.79). That interval contains the RQ2 bound of about one point per tenfold increase (Section 3.6). On this scale, then, the adjusted rise cannot be told apart from the smallest effect of interest. The random-band control gives higher net estimates (Appendix F.4). Figure 1 shows the measured share and the share matched by chance in each quintile of size.

\begin{figure}[htbp]\centering
\includegraphics[width=0.96\linewidth,height=0.82\textheight,keepaspectratio]{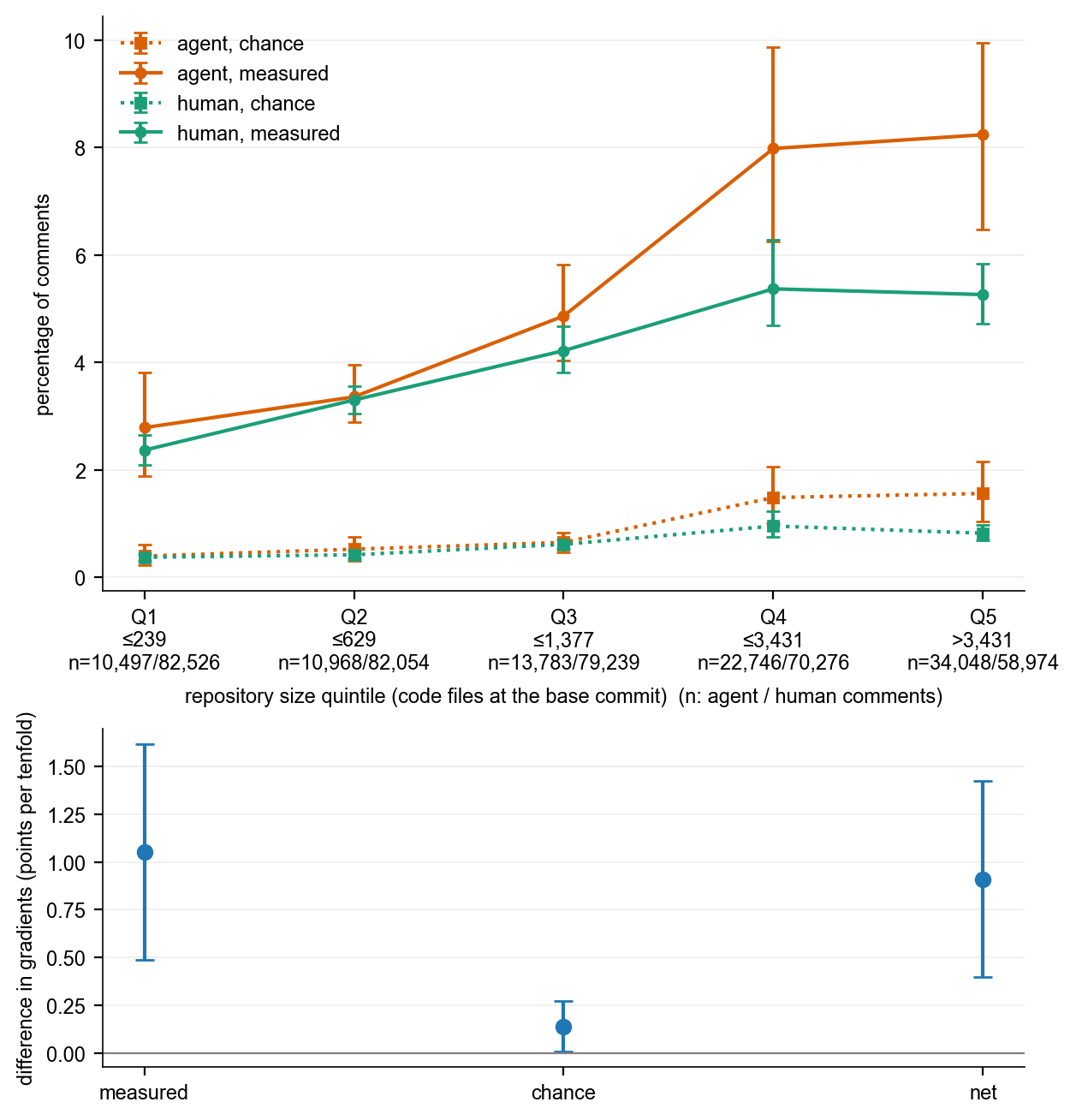}
\caption{Top: share of review comments that name a repository item outside the change, by quintile of repository size (code files at the base commit; the quintiles are cut once over all plotted comments of both arms pooled, and each tick gives the quintile's upper edge and the number of agent and human comments) and arm (solid), with the share the same names match by chance in a repository of similar size (dotted); all eligible comments not addressed to an agent, both arms; 95 per cent intervals from a bootstrap over repositories. Bottom: in the RQ3 comparison set, the difference between the arms' size gradients on the percentage-point scale, measured, by chance and net of chance, with repository fixed effects.}
\end{figure}
Exploratory analyses of the curve's shape, added after the results (D22), find the rise concentrated between small and mid-sized repositories (as Section 1 defines them). In Figure 1, the agent share climbs from the first to the fourth quintile of size. It is flat between the fourth quintile, whose upper edge is about 3,400 code files, and the fifth. A term for the square of log size, added to the RQ2 model, is negative: -0.21 on the log-odds scale (interval -0.39 to -0.04). It is negative in the human arm too (-0.12, interval -0.20 to -0.04). Fitted separately below 1,500 code files, the RQ2 model gives an odds ratio of 1.67 (interval 1.14 to 2.43). At or above that size it gives 0.97 (interval 0.67 to 1.41), an imprecise estimate rather than evidence of a plateau. We chose this split after seeing the quintiles, so the quadratic term is the evidence that does not depend on a chosen split. On repositories with at least 300 code files, the odds ratio is 1.24 (interval 0.99 to 1.54). Appendix B lists these models.

The rise holds when the unit, the reviewers or the repository's other properties change. Figure 2 shows the agent-arm size odds ratio under each such change, and Appendix B describes each specification. The estimated gradients vary across agent groups (Appendix E). Two components of the measured association both increase with size on average: reviewers in larger repositories name the project more often, and what they name more often lies outside the change (Appendix E).

\begin{figure}[htbp]\centering
\includegraphics[width=0.96\linewidth,height=0.82\textheight,keepaspectratio]{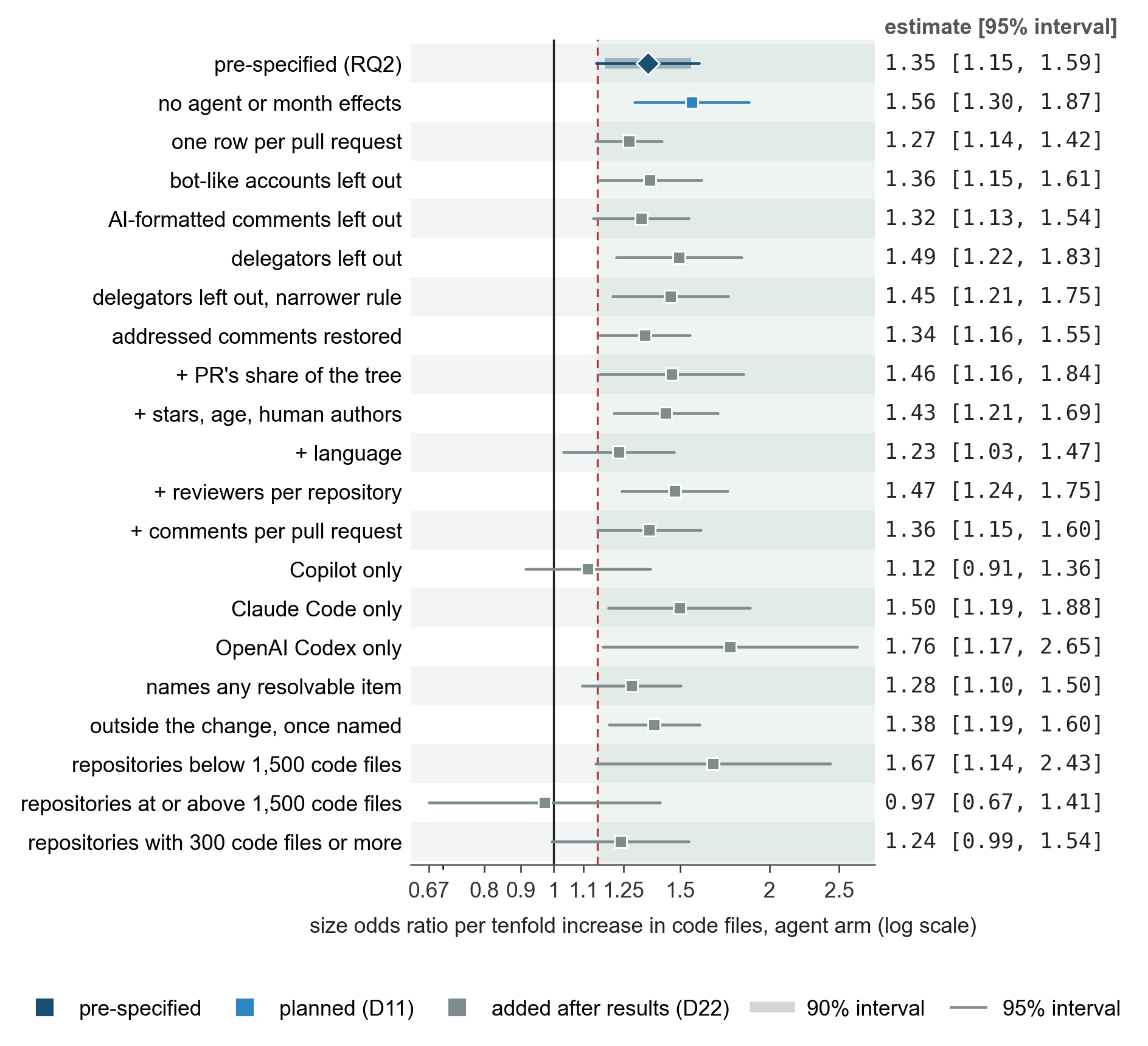}
\caption{The agent-arm size odds ratio per tenfold increase in code files under the pre-specified model and every specification and population sensitivity, including the fits below and at or above 1,500 code files and at 300 code files or more, with 95 per cent intervals (thin lines) and, for the pre-specified model, the 90 per cent interval of its equivalence test (thick bar), on a log scale. The dashed line marks RQ2's smallest effect of interest, 1.15, and the green region lies above it; the grey bands on alternate rows only guide the eye; colour marks whether a specification was pre-specified, planned (D11) or added after the results (D22). The error-corrected estimate, which uses a simpler model, is reported in Section 4.4.}
\end{figure}
\subsection*{4.3 Agent and human gradients within the same repositories (RQ3)}
\addcontentsline{toc}{subsection}{4.3 Agent and human gradients within the same repositories (RQ3)}
Comments on human pull requests in the same repositories show a similar rise with size. The pre-specified comparison finds the agents' rise slightly steeper, within the bound for a negligible difference. The repository-month comparison finds no detectable excess as a log-linear average (equivalent to 1 within the bound). With a square of size per arm, both comparisons find a ratio above 1 at the mean size, with the estimated excess concentrated among small and mid-sized repositories.

\textbf{Each arm.} With reviewer insider status and month effects, the size odds ratio in the comparison set is 1.41 for human pull requests (interval 1.32 to 1.51). For agent pull requests it is 1.40 (interval 1.19 to 1.64). Appendix E gives each arm's gradient raw and adjusted, on both scales.

\textbf{The pre-specified comparison.} This model compares the two rises while adjusting for each repository's means of its terms (the Mundlak device, Section 3.6), pooled over agents and months. The ratio of the agent to the human size odds ratio is 1.15 (interval 1.04 to 1.27). The Holm-adjusted p value is 0.005. The equivalence test within 1/\allowbreak{}1.25 and 1.25 gives p = 0.0484. The upper end of its 90 per cent interval, 1.2490, sits on the bound. The pre-specified verdict is therefore ``different from zero, and smaller than the smallest effect of interest''. Net of chance, with repository fixed effects, the agent gradient exceeds the human gradient by 0.91 percentage points per tenfold increase (interval 0.40 to 1.42). This is about the size of the bound on that scale; Section 6.2 explains why the arms' slopes on that scale are not directly comparable. The lower panel of Figure 1 shows the difference measured, by chance and net of chance.

\textbf{The repository-month comparison.} Added after the results, this comparison holds the repository, the month, pull-request size and the agent group fixed (Section 3.7). Its ratio is 1.06 (interval 0.92 to 1.22), equivalent to 1 within the bound. Without the delegators' comments, its verdict is inconclusive (Table 2). Size barely varies within a repository-month cell, so the size term of this comparison, like that of the pre-specified one, is identified mainly from differences between repositories. What the cells remove is the mix of agents, months and pull-request sizes.

\textbf{Other specifications.} The result depends on the specification and on how size is modelled. Adding agent-group and month effects to the pre-specified model (added after the results) removes the excess (Table 2). Across the other specifications, the excess is absent in some and near or beyond the bound in others (Table B2, Figure 3). With a square of size per arm (added after the results), both comparisons find a ratio above 1 at the mean size, 1.23 and 1.22 (Table 2). The excess lies among small and mid-sized repositories, and the estimated ratio falls with size; Appendix B gives the size-specific estimates. A slope per size class detects an excess only nominally, among small repositories (Table B4). Appendix B reports these models, the comparisons with each agent group and with human authors of different standing, and the delegator sensitivities.

\needspace{23\baselineskip}
{\footnotesize
\setlength{\tabcolsep}{4pt}
\begin{tabularx}{\linewidth}{Xl>{\raggedleft\arraybackslash}l>{\raggedleft\arraybackslash}l>{\raggedleft\arraybackslash}l>{\raggedleft\arraybackslash}lX}
\caption{The arm comparison under the specifications this section discusses: the ratio of the agent to the human size odds ratio, its 95 per cent interval, the 90 per cent interval that decides the equivalence test, and the pre-specified decision rule applied to it (equivalence bound 1.25; Holm adjustment for the pre-specified model only). Status: pre-specified or added after the results (D22). Rows with a square of size per arm give the ratio at the mean log size over the analysis frame, about 845 code files (Section 3.4). For models with a square of size per arm, the ratio is the exponentiated difference in the arms' local log-odds slopes at the stated size; it is not a ratio of tenfold-change odds ratios. Rows that do not name the same repository and month modify the pre-specified model. Upper ends of the 90 per cent interval are given to four decimals. Table B2 in Appendix B and Figure 3 give every specification. † marks a TOST p value between 0.045 and 0.055, on the decision boundary. Appendix F.4 notes the pre-specified model on the fixed-effects sample.}\\
\toprule
\textbf{Specification} & \textbf{Status} & \textbf{Ratio} & \textbf{95 per cent interval} & \textbf{90 per cent interval} & \textbf{TOST p} & \textbf{Verdict} \\
\midrule\endfirsthead
\toprule \textbf{Specification} & \textbf{Status} & \textbf{Ratio} & \textbf{95 per cent interval} & \textbf{90 per cent interval} & \textbf{TOST p} & \textbf{Verdict} \\ \midrule\endhead
pre-specified (Mundlak) † & pre-specified & 1.15 & 1.04 to 1.27 & 1.06 to 1.2490 & 0.0484 & different from zero, and smaller than the smallest effect of interest \\
same repository and month + PR size + agent group & D22 & 1.06 & 0.92 to 1.22 & 0.94 to 1.1936 & 0.0112 & equivalent to zero within the smallest effect of interest \\
same repository and month + PR size + agent group, delegators' comments left out & D22 & 1.11 & 0.94 to 1.31 & 0.96 to 1.2772 & 0.0819 & inconclusive: neither different from zero nor within the smallest effect of interest \\
pre-specified + agent group and month & D22 & 0.99 & 0.91 to 1.09 & 0.92 to 1.0707 & < 0.0001 & equivalent to zero within the smallest effect of interest \\
pre-specified + a square of size per arm (ratio at the mean log size) & D22 & 1.23 & 1.07 to 1.42 & 1.09 to 1.3919 & 0.4221 & different from zero \\
same repository and month + PR size + agent group + a square of size per arm (ratio at the mean log size) & D22 & 1.22 & 1.04 to 1.43 & 1.07 to 1.3947 & 0.3817 & different from zero \\
\bottomrule
\end{tabularx}}
Figure 3 shows the specifications of Table B2; Appendix F.4 notes the pre-specified model on the fixed-effects sample.

\begin{figure}[htbp]\centering
\includegraphics[width=0.96\linewidth,height=0.82\textheight,keepaspectratio]{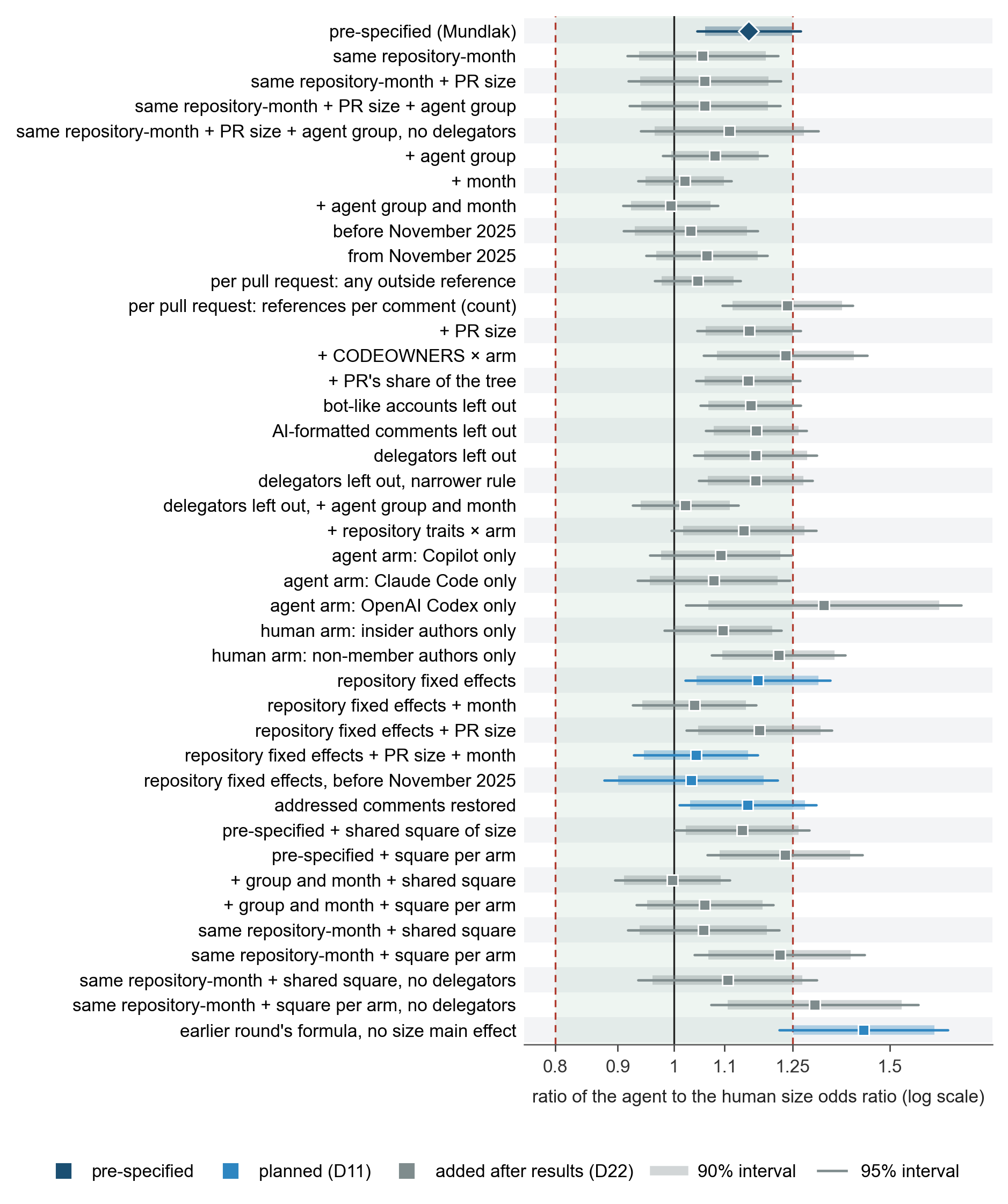}
\caption{The arm comparison of Table B2: the ratio of the agent to the human size odds ratio under every specification, with its 95 per cent interval (thin line) and the 90 per cent interval that decides the equivalence test (thick line), on a log scale. The dashed lines mark the equivalence bounds, 1/\allowbreak{}1.25 and 1.25, and the green region lies between them; the grey bands on alternate rows only guide the eye; rows that do not name another base model (the same repository and month, repository fixed effects or the earlier round's formula) modify the pre-specified model; the eight rows before the last add a square of size, shared or per arm, the per-arm rows read at the mean log size, about 845 code files (Table B2; Table B3 reads the per-arm models at three sizes and Table B4 gives a slope per size class); colour marks whether a specification was pre-specified, planned (D11) or added after the results (D22). Appendix F.4 notes the pre-specified model on the fixed-effects sample.}
\end{figure}
The pre-specified comparison finds a small agent excess at the edge of the equivalence bound. A comparison within the same repository and month finds no detectable excess as a log-linear average (equivalent to 1 within the bound). Across specifications and size models, the evidence does not establish a robust agent-specific excess.

\subsection*{4.4 Measurement}
\addcontentsline{toc}{subsection}{4.4 Measurement}
The instrument is usually right when it finds a reference, but it misses many. On the stratified validation sample of 600 comments (Section 3.8), its weighted precision is 79.7 per cent overall, 81.6 per cent in the agent arm and 78.8 per cent in the human arm.

Its recall is 43.6 per cent overall, 59.3 per cent in the agent arm and 38.8 per cent in the human arm. These figures rest on 21 misses: 18 from people's labels and 3 from model judgements confirmed by the index. Under a second rule, added after the results, that also counts the 1 judged miss that the index does not confirm, recall is 40.4 per cent (interval 27.3 to 61.4). The false-omission rate is 4.7 per cent in the agent arm and 5.4 per cent in the human arm. The sample holds too few misses to show how recall changes with size. Precision falls with size in the human arm and is flat in the agent arm. Appendix C gives both by size third.

People's checks support the labels. People checked 186 of the validation's labels and agreed with the final model label on 95 per cent. Appendix C gives the planted reversals and the agreement among judges and readers.

The files reviewers name are later changed in the same pull request far more often than other files. We took the 18,061 flagged comments whose named files lay outside the change when the comment was written. Of the named files, 10.9 per cent are among the files the pull request changes by its end. A named file is changed 11.7 times as often as a file drawn from its own directory (interval 9.8 to 14.2). Appendix C gives the baselines.

In the judges' labels, the comment directs the contributor to the item, to look at, reuse, follow or update it, in 76 per cent of the confirmed references. The share is 68 per cent on agent and 79 per cent on human pull requests. It rests on 233 confirmed references, so the difference between the arms is indicative.

The known-groups hypothesis is not supported. In the agent arm, insider reviewers name outside items no more often than other reviewers, and the odds ratio of 0.89 (interval 0.68 to 1.17) fails the plan's at-or-below-1 rule. In the human arm, the odds ratio is 1.00 (interval 0.90 to 1.12), above 1 by a negligible margin. Neither result supports the hypothesis that insider reviewers name outside items more often than other reviewers. Appendix F.1 gives the check within repositories and without the delegators.

Corrected for the instrument's measured error, the estimates are imprecise. With design-based supervised learning (Section 3.9), the size odds ratio is 1.22 (interval 0.88 to 1.69), against 1.56 uncorrected in the same specification. The ratio of the agent to the human gradient is 0.80 (interval 0.32 to 2.00), against 1.07 uncorrected (interval 0.92 to 1.26). With all four Mundlak means of the pre-specified model, the corrected ratio is 1.36 (interval 0.50 to 3.72). Neither corrected interval excludes 1. The corrections rest on 300 and 493 labelled comments.

\subsection*{4.5 Context files (RQ4, exploratory)}
\addcontentsline{toc}{subsection}{4.5 Context files (RQ4, exploratory)}
We find no clearly estimated association between context-file presence and the size gradient. The analysis uses 85,183 comments on agent pull requests other than Devin's. The ratio of the size odds ratio with the agent's context file present at the base commit to the size odds ratio without it is 1.04 (interval 0.78 to 1.39). The interval admits a flattening or a steepening of a fifth or more. Appendix B gives the sensitivities.

\section*{5 Discussion}
\addcontentsline{toc}{section}{5 Discussion}
\subsection*{5.1 What the rise means}
\addcontentsline{toc}{subsection}{5.1 What the rise means}
Three findings describe how outside references change with size, and they differ in strength. First, on average, a log-linear model finds an association. On agent pull requests, the odds that a comment names an existing item outside the change rise by an odds ratio of 1.35 per tenfold increase in code files (interval 1.15 to 1.59). Net of what the same names match by chance in a repository of similar size, the share of such comments still rises with size. Second, exploratory analyses of the curve's shape place the rise mainly between small and mid-sized repositories. Third, among large repositories the estimate is imprecise, so the data neither show nor rule out a continued rise (Section 4.2). The average association remains when we hold fixed, or leave out, the pull request's coverage of the tree, comments from bots or delegators, and the repository's stars, age and number of authors (Figure 2; Appendices F.1 and F.2).

Files that reviewers name are changed later in the same pull request more often than other files: 11.7 times as often as another file of the same directory. In the judges' labels, most confirmed references direct the contributor to the item. Neither result shows that the contributor acted on the comment, and the unsupported known-groups hypothesis means that naming an item need not reflect knowledge that only some reviewers hold (Section 6.1).

Two accounts fit the rise. The simpler one is supply: a larger repository holds more existing helpers, modules and conventions, so reviewers have more to name, whatever the contributor saw. It is consistent with a shared rise, with a small, specification-dependent arm difference, and is silent on the known-groups check. The second account, which motivated the study, is a bounded view. A contributor works from a bounded part of the repository. In a larger repository that part covers less of the whole, and reviewers then name what lies beyond it, such as an existing helper to call or a module the change should follow. This account adds a prediction that supply does not make: contributors who know the project less should receive more pointers. The human gradient barely differs between insider and non-member authors (Appendix B), which does not favour the bounded view, though that contrast is weak (Appendix F.2).

The shape of the rise does not separate the two accounts. A flattening among large repositories fits supply if what reviewers find worth naming saturates once a repository holds enough of it. It fits the bounded view if reviewers' own view of very large repositories is also bounded. It also fits a measure whose recall falls in the largest trees (Section 6.2). Prior work supplies the premises of the bounded-view account. Many coding conventions, such as naming, are unwritten and emerge within a codebase, and inferring them requires a global view of it (Allamanis et al., 2014). The context into which a change is plugged is one of reviewers' information needs (Pascarella et al., 2018), and reviewers share links to meet such needs (Wang et al., 2021). Review comments raise duplicated functionality and other symptoms of architecture erosion, and most of those raised are addressed (Li et al., 2022). None of these studies relates what reviewers bring from the wider system to the size of that system. We add that dimension.

\subsection*{5.2 Agent and human pull requests}
\addcontentsline{toc}{subsection}{5.2 Agent and human pull requests}
The bounded-view account predicts a steeper rise for agent than for human pull requests. In one Python repository, agents often miss chances to reuse existing code (Huang et al., 2026). Claude Code pull requests that needed revision were often revised for adherence to project-specific standards (Watanabe et al., 2026). If an agent sees less of a large repository than a person who works in it, reviewers should point it outside the change more often. How well agents find the relevant code is itself open; current systems add search, localisation and dependency-analysis steps for it (Bairi et al., 2024; Yang et al., 2024; Xia et al., 2025; Section 2.4). The premise also concerns the agent's own view. Over half of the agent-arm comments are on pull requests that a developer opened under their own account (Section 3.2), and that developer may supply what the agent did not see.

Our data give this prediction weak and specification-dependent support. Human pull requests in the same repositories also show a rise with size. The pre-specified comparison finds the agents' rise steeper by a ratio of 1.15 (interval 1.04 to 1.27), within the bound for a negligible difference, but only just. It is within the bound only before considering the biases discussed in Section 6.3. The repository-month comparison finds no detectable excess as a log-linear average: its ratio is 1.06 (interval 0.92 to 1.22), equivalent to 1 within the bound. Under both curve forms, the agents' excess is largest among small repositories and not detected among large ones, a pattern that does not support the premise where it should apply most (Appendix B). On the percentage-point scale, the pre-specified companion puts the excess at about the size of the bound. It narrows when the month is held fixed, and the arms' recall differs, so that scale cannot settle the question either (Appendix E).

The plan reads a ratio within the bound as a gradient that belongs to large codebases for everyone. Similar rises with size, however, do not settle every difference between agent and human pull requests. An analysis of integration friction finds agent pull requests more sensitive to the repository than human ones, with codebase size removing little of that (Russo, 2026). Repository properties other than size may therefore act on agent pull requests more than on human ones. Equal rates per comment also leave cumulative cost open. Whether review guidance persists across subsequent agent tasks depends on how it is retained and reused.

\subsection*{5.3 Implications for practice and for tool builders}
\addcontentsline{toc}{subsection}{5.3 Implications for practice and for tool builders}
\textbf{Maintainers.} In exploratory analyses, review in mid-sized repositories includes more pointers to existing code than review in small ones, in both studied arms. Appendix E gives the predicted shares at typical sizes. The items named belong to the project, not to the contributor. In the judges' labels, most comments that name them direct the contributor to the item; the apparent difference between arms is indicative only (Section 4.4). Directive items that recur are candidates to be written down once and checked by tools. Rules that hold across a codebase, such as layering and allowed dependencies, can be encoded for static conformance checking (Passos et al., 2010). Practitioners, however, often check architecture consistency by informal means, partly because mapping a system to its architecture takes effort (Ali et al., 2018).

\textbf{Agent context.} A common proposal for agents is more repository context, such as a context file that the agent reads (Gloaguen et al., 2026). On one benchmark, retrieval baselines resolved fewer issues as the retrieved context grew longer (Jimenez et al., 2024). For the models tested there, more retrieved context did not by itself help. Our exploratory RQ4 finds no clearly estimated association between context-file presence and the size gradient (Section 4.5). Projects that write such files may differ from those that do not, so the question needs a before-and-after comparison within repositories that add a file. Context files most often cover testing practice, conventions, implementation details and architecture (Chatlatanagulchai et al., 2026; Mohsenimofidi et al., 2026). Most comments with an outside reference, by contrast, name a defined symbol (Section 4.1). Retrieving repository code at the time of the change improves completion over an in-file baseline (Zhang et al., 2023), and it is a hypothesis for what would serve those references.

\textbf{Review tooling.} A project can apply the resolution rules of Section 3.4 to its own review history and list the items its reviewers name most often outside changes. The rules detect candidate references, not intent. Common library names are their recurring false positive and would rise in such a list. The negative controls give each name's chance-match rate, which can serve as a stop list. The rules cannot tell a comment that directs the contributor from one that names an item in passing, so the list needs reading. The items that reviewers ask contributors to use or follow are candidates for a context file, for an agent's retrieval, or for a check that flags a new definition close to an existing one. Comments from language-model review tools that are concise, contain code and are manually triggered, particularly those from hunk-level tools, are more likely to lead to changes (Sun et al., 2026). That is the form an automated pointer to an existing item would take.

\subsection*{5.4 Implications for research}
\addcontentsline{toc}{subsection}{5.4 Implications for research}
\textbf{Chance-match controls for text-to-code linking.} Any measure that resolves names against a repository, and is compared across repository sizes, carries a size gradient of its own, because a larger tree offers more names to match by chance. Studies that link developer text to code by name report precision per system and do not relate chance matches to the size of the codebase (Bacchelli et al., 2010; Dagenais and Robillard, 2012; Rigby and Robillard, 2013). In our data, chance matching also grew with size, though more slowly than the measured share (Section 4.2). A size-matched negative control, run with the same instrument, estimates this gradient at little extra cost.

\textbf{Validation that can see size.} If the instrument's recall changed with size, that change would enter the gradient directly. Our validation sample holds too few misses to show it. What this question needs next is a human-labelled reference, stratified by arm and size third, with enough misses in each cell; Appendix C gives the sample sizes this would take.

\textbf{Known groups for reference measures.} Repository association is a coarse marker of knowledge. Many developers review changes to a module they never authored, and the share of reviewers without expertise in a module relates to its post-release defects (Thongtanunam et al., 2016). A known-groups check (Cronbach and Meehl, 1955) built on review-aware ownership of the named item would likely give a sharper contrast for measures of this kind.

\textbf{Design ideas.} A count of named outside items can enter studies of review themes (Haider and Zimmermann, 2026) as a variable beside their themes. Coding the comments by what the reviewer asks for (reuse, a convention, placement or a passing mention) would give the count its reasons. We did not code them here: coding by function needs trained human coders, and the validation's judges answered only whether a comment directs the contributor to the item. Two further designs would show whether reviewers point agents to the same code again where people learn it: comparing agents with first-time contributors, and following a contributor's successive pull requests in one repository.

\section*{6 Threats to validity}
\addcontentsline{toc}{section}{6 Threats to validity}
This section states each limitation once and what it does to the reading of the results. Appendix F gives every threat in detail with the checks behind it.

\subsection*{6.1 Construct validity}
\addcontentsline{toc}{subsection}{6.1 Construct validity}
The outcome records whether a comment names an existing repository item outside the change. It does not measure why the reviewer named it, whether the contributor acted on the comment, or whether the reference reflects knowledge of the project. Named files are later changed more often than comparable files. This shows that the comment and a later change concern the same file, not that the comment led to it. The check also bears on comments that ask for the named file to change, not on pointers to reuse, which are followed by calling the item from the contributor's own files. The known-groups hypothesis is not supported: in the agent arm the odds ratio fails the plan's at-or-below-1 rule, and in the human arm it is above 1 by a negligible margin, so neither arm shows insider reviewers naming outside items more often than other reviewers. It therefore gives no support to reading a named item as a transfer of project knowledge. Insider status comes from GitHub's author\_\allowbreak{}association field as read at collection, which blurs insiders and others (Appendix F.1).

\subsection*{6.2 Measurement error}
\addcontentsline{toc}{subsection}{6.2 Measurement error}
The instrument is usually right when it flags a comment but misses many references (Section 4.4). The reported shares therefore likely understate how often reviewers name outside items. The central measurement threat is undercounting that differs by arm or changes with repository size, and the validation holds too few misses to test it. The instrument's uniqueness limits should lower recall in larger trees, which would pull the gradients toward 1. Precision, however, falls with size in the human arm and is flat in the agent arm, which adds to the human gradient. The error-corrected size estimate is also lower than the uncorrected one in the same specification. The net direction of the measurement error is therefore unknown. Corrected for the measured error, in simpler specifications than the pre-specified ones, the estimates have wide intervals that include no effect, so they qualify the measured estimates rather than replace them. The validation's reference rests partly on model judgements, which people checked only in part. On the percentage-point scale, the lower recall on human pull requests shrinks the human slope, so the arms' slopes on that scale, including the net-of-chance companion of Section 4.3, cannot be compared directly (Appendix E).

\subsection*{6.3 Internal validity of the agent-human comparison}
\addcontentsline{toc}{subsection}{6.3 Internal validity of the agent-human comparison}
Agent and human pull requests are not randomly assigned, so the comparison estimates associations, not effects of agent authorship. The arms differ in agent and month composition, pull-request size and who comments. Holding the month fixed removes the excess but may also remove part of an agent effect (Appendix F.2). The delegator rule is broad: it also marks reviewers who were only asked to review (Section 3.3). Some human-arm pull requests may have been written with an agent, which would draw the two gradients together.

The agent-specific excess depends on the specification and on how size is modelled (Section 4.3, Appendix B). With a quadratic in log size per arm, both comparisons find the agents' rise steeper among small and mid-sized repositories, unless agent-group and month effects are added to the pre-specified model (Table B3). With a slope per size class, the excess appears only among small repositories, within the same repository and month. That result is nominal: significant before but not after Holm adjustment over the three size classes, and only with the delegators' comments kept (Table B4).

The Mundlak device does not remove every repository characteristic, and both comparisons rest mainly on differences between repositories. The size gradient itself compares repositories with one another, not one repository as it grows. The repository traits we measured do not explain it (Appendix F.2).

The biases identified here, agent-written pull requests in the human arm and falling precision that adds more to the human gradient, push the arm ratio toward 1. They therefore favour the ``smaller than the smallest effect of interest'' half of the pre-specified verdict. A rough calculation suggests that correcting for the observed precision gradient would by itself move the upper end of the 90 per cent interval above 1.25 (Appendix F.1). The precision estimates by size third, however, are themselves uncertain.

\subsection*{6.4 External validity}
\addcontentsline{toc}{subsection}{6.4 External validity}
The frame consists of public GitHub repositories with at least 100 stars in which six agent groups opened pull requests between December 2024 and September 2026. The measure covers only the inline comments that open review threads. The results may not generalise to smaller or private projects, other platforms, other agents, or the other channels through which reviewers and developers talk to agents.

\subsection*{6.5 Conclusion validity}
\addcontentsline{toc}{subsection}{6.5 Conclusion validity}
The analysis plan was frozen internally after estimates from an earlier round had been seen, and was not publicly registered. Several sensitivity analyses were added after the full-frame estimates were seen (D19, D20 and D22; Section 3.11). They show how the answers depend on the specification and do not replace the pre-specified verdicts. The arm comparison's verdict sits on the decision boundary (Table 2). It changes with the estimator, the sample and the functional form of size (Appendix F.4). Power and error rates were simulated for the earlier round's frame and estimators, not for this one.

\section*{7 Conclusion}
\addcontentsline{toc}{section}{7 Conclusion}
Review comments on agent and human pull requests name existing code beyond the change more often in larger repositories on average, and exploratory analyses place the rise mainly between small and mid-sized repositories. On agent pull requests, 6.5 per cent of eligible comments name such an item, and the rise remains net of chance matches as size-matched negative controls estimate them. Human pull requests show a similar pattern. A pre-specified comparison finds a small agent excess just within the bound set as negligible, but the evidence does not establish a robust agent-specific excess.

These are associations measured by an instrument that misses many references (Section 6).

The study contributes a reproducible measure of outside-change references, size-matched negative controls, and a within-repository comparison of agent and human pull requests. Future work needs validation that can estimate recall across repository sizes and designs that separate agents from contributors new to a repository. For maintainers and tool builders, the items reviewers name repeatedly, above all those they ask contributors to use or follow, are candidates for retrieved context and automated checks.

\section*{Data availability}
\addcontentsline{toc}{section}{Data availability}
The study's data and code are deposited at Zenodo (\url{https://doi.org/10.5281/zenodo.23290164}) and mirrored on GitHub (\url{https://github.com/Gatem/beyond-the-change-replication}); the analysis tables are also on Hugging Face (\url{https://huggingface.co/datasets/SabryFarrag/review-pointers-beyond-the-change}). The deposit holds the list of repositories in the frame; the review-comment table, with the identifiers that retrieve each comment and its pull request from GitHub; the measured links and negative controls; the analysis table; and the validation sample with every judgement. Standalone scripts with their configuration recompute the estimates, and the numbers that rest on them, from the analysis table. The steps that need the repositories themselves (the measurement, the negative controls, the completeness check, the convergent check and review coverage) are deposited as outputs, and a second route of the scripts collects the data again from GitHub and reruns them. The validation's labels and scores are deposited and reused, not recomputed. Review comments are public GitHub content; the deposit keeps their identifiers so that each can be retrieved from its source.

\section*{Declarations}
\addcontentsline{toc}{section}{Declarations}
\textbf{Funding.} No funding was received for this study.

\textbf{Competing interests.} The author has no competing interests to declare.

\textbf{Ethics approval.} Not applicable. The study analyses public GitHub data. For the volunteers who checked labels, only each reader's rate of catching planted reversals was recorded, to decide whether their labels were used; the deposit pseudonymises the readers.

\textbf{Consent to participate.} The volunteers who checked labels agreed to take part.

\textbf{Consent for publication.} Not applicable.

\textbf{Code availability.} The analysis scripts are part of the deposit described under Data availability.

\textbf{Author contributions.} The study has a single author, who conceived it, carried it out and wrote the manuscript.

\textbf{Use of generative AI.} Generative AI tools were used in preparing this manuscript, including drafting text and analysis code. The author reviewed all content and takes full responsibility for it.

\clearpage
\setcounter{table}{0}\renewcommand{\thetable}{A\arabic{table}}
\setcounter{figure}{0}\renewcommand{\thefigure}{A\arabic{figure}}
\section*{Appendix A Deviation log and pilot gate}
\addcontentsline{toc}{section}{Appendix A Deviation log and pilot gate}
{\footnotesize
\setlength{\tabcolsep}{4pt}
\begin{tabularx}{\linewidth}{X>{\raggedleft\arraybackslash}lXX}
\caption{Departures from the pre-specified plan, with the date each was decided. D1 to D18 were decided before the present run's full-frame outcome had been computed, D19 to D22 after; the four rows dated 2026-10-11 record later changes under D22.}\\
\toprule
\textbf{} & \textbf{Date} & \textbf{What changed} & \textbf{Why} \\
\midrule\endfirsthead
\toprule \textbf{} & \textbf{Date} & \textbf{What changed} & \textbf{Why} \\ \midrule\endhead
D1 & 2026-10-08 & Frame enlarged to 4,138 repositories (Section 3.1) & Power for RQ2 \\
D2 & 2026-10-08 & 31 capped listings completed by date-sliced search & Complete listings \\
D3 & 2026-10-08 & Collection, measurement and analysis rebuilt as configured steps & One implementation \\
D4 & 2026-10-08 & Symbol definitions restricted to files present at the base commit & Items that existed \\
D5 & 2026-10-08 & The change judged at the comment's commit, renames off & Change as the reviewer saw it \\
D6 & 2026-10-08 & Size counted from the index at each base commit & Size when the PR began \\
D7 & 2026-10-08 & Random-band control added beside the next partner & Second chance estimate \\
D8 & 2026-10-08 & Net contrasts as separate and stacked fits, identity checked & Shared effects in the old fit \\
D9 & 2026-10-08 & Size main effect added to the RQ3 model; old formula kept as a sensitivity & Correct specification \\
D10 & 2026-10-09 & Pre-specified estimators restored as the confirmatory models & Restores the plan \\
D11 & 2026-10-09 & Planned sensitivities and known-groups models added (Sections 3.7 and 3.8; Appendix B) & Robustness and validity \\
D12 & 2026-10-09 & Single-rater precision check replaced by the stratified validation (Section 3.8) & Precision and recall \\
D13 & 2026-10-09 & Error correction (DSL) added as a sensitivity & Instrument error \\
D14 & 2026-10-09 & Agent signatures matched case-sensitively, as in the plan's script & No PR changed arm \\
D15 & 2026-10-09 & Pilot gate and results review added & Checks before the run \\
D16 & 2026-10-09 & Convergent check run, against random files of the same tree & Not run before \\
D17 & 2026-10-09 & RQ4 context file counted by presence at the base commit; Jules on AGENTS.md & Data available \\
D18 & 2026-10-10 & Random band widened stepwise to 30, 50 and 100 per cent of size & Partners for rare languages \\
D19 & 2026-10-10 & Band contrast for RQ2, known groups within repositories, precision and recall by size, people's check & Language-model review of the full results \\
D20 & 2026-10-10 & Next partner re-paired on code files at the head & Language-model review, round 3 \\
D21 & 2026-10-10 & The change limited to the PR's own files; every comment measured again & Restores the plan's definition \\
D22 & 2026-10-10 & The sensitivities listed under D22 in Appendix B; decision rule applied to all; second recall rule & Language-model review of the first draft \\
D22 & 2026-10-11 & Repository-month models re-specified with the size main effect kept as a term; earlier value 1.09 (interval 0.94 to 1.26) & Language-model review, round 7 \\
D22 & 2026-10-11 & Functional form of the RQ2 gradient checked (quadratic term, split at 1,500 code files, 300 code files and above) & Language-model review, round 8 \\
D22 & 2026-10-11 & Arm comparisons refitted with a square of size, shared and per arm, and read at 300, 1,000 and 3,000 code files (Table B3) & Language-model review, round 9 \\
D22 & 2026-10-11 & Arm comparisons refitted with a linear spline in log size per arm, knots at 300 and 3,000 code files (Table B4) & Language-model review, round 10 \\
\bottomrule
\end{tabularx}}
\textbf{Timing of the departures.} Before the full-frame estimates were computed, we enlarged the frame, completed capped repository listings, revised how the changed files and repository size are measured, restored the planned estimators and added the pilot gate (D1 to D18). For all but D1, these decisions came after the earlier round's full-frame estimates had been seen. D10, D14 and D17 came after the pilot's first results, and D18 after the pilot had passed the gate. D18 widened the random band when the band control's own check failed on the full frame, before any full-frame estimate was computed. It changed only that control's configuration, which is not among the files the lock records, so the lock was not renewed for it. After the full-frame estimates were seen, we added sensitivity, validation and robustness analyses in response to language-model reviews (D19, D20, D22). These additions change no pre-specified model or verdict. The four rows dated 2026-10-11 record later changes made in response to reviews: the re-specified repository-month models, the checks of the gradient's functional form, and the arm comparisons refitted with a curved size gradient, as a quadratic (Table B3) and as a linear spline (Table B4). D21 is not an addition: it limits the change to the files the pull request itself changes, as Section 3.4 defines it, and measures every comment again; the pre-specified verdicts are the same before and after it.

\textbf{The pilot gate.} The pilot of Section 3.10 covered 60 seeded repositories. Since partners could come only from those 60, the pilot used random-band partners within 60 per cent of size without matching language, and drew 30 validation items per stratum. The pilot gate required:

\begin{itemize}
\item (P1) the pilot's scope;
\item (P2) an independent validation of at least 100 items with a fact-check failure rate of at most 5 per cent and an arm difference in that rate of at most 5 percentage points;
\item (P3) two implementations of every model and contrast;
\item (P4) the identity net = measured minus chance on unrounded values;
\item (P5) the model hierarchy;
\item (P6) completeness, with a shortfall of at most 1 per cent and capped repositories listed;
\item (P7) a review by model families other than Claude with every finding triaged;
\item (P8) a lock recording the hashes of the instrument, the estimators and the configurations of the frame, the measurement, the next-partner control and the estimates.
\end{itemize}
The pilot first passed the gate on 2026-10-09. We then changed the estimator and frame configurations to restore the pre-specified estimators, add the decision tests and add RQ4. We ran the pilot again on the changed files, repeated its review by other model families and triaged every new finding. It passed the gate at 18:29 UTC on 2026-10-09, which re-froze the lock before any full-frame outcome had been computed.

\textbf{Results review.} As quality control before the results were written, two reviewers from model families other than Claude, gemma4:12b (Google) and qwen3:14b (Alibaba), reviewed the model configuration, the full-frame estimates, the error-corrected estimates, the validation, the sample flow and the completeness check, and we gave every finding a verdict with its reason. These reviews, and the reviews named under D19 to D22, were carried out by language models.

\setcounter{table}{0}\renewcommand{\thetable}{B\arabic{table}}
\setcounter{figure}{0}\renewcommand{\thefigure}{B\arabic{figure}}
\section*{Appendix B Sensitivity catalogue}
\addcontentsline{toc}{section}{Appendix B Sensitivity catalogue}
The planned sensitivities (D11):

\begin{itemize}
\item rq2: RQ2 without the agent and month fixed effects;
\item rq3: RQ3 with repository fixed effects in place of the Mundlak device;
\item rq3\_\allowbreak{}frozen: the formula of the earlier round's RQ3 script, with repository fixed effects and without the size main effect;
\item rq3\_\allowbreak{}with\_\allowbreak{}addressed: RQ3 with comments addressed to an agent restored;
\item rq3\_\allowbreak{}size\_\allowbreak{}time: RQ3 with PR size and repository and month fixed effects;
\item rq3\_\allowbreak{}no\_\allowbreak{}truncated: RQ3 without repositories whose listing is still capped; after the 31 capped repositories are completed none is, so this model reproduces the comparison sample and serves as a completeness check, not a robustness check;
\item rq3\_\allowbreak{}aidev\_\allowbreak{}window: RQ3 on pull requests created before 2025-11-01;
\item rq3\_\allowbreak{}net\_\allowbreak{}band: the net-of-chance contrast with the random-band control;
\item rq3\_\allowbreak{}net\_\allowbreak{}band15: the same contrast on repositories matched within the pre-specified 15 per cent of size, fitted as rq3\_\allowbreak{}lpm\_\allowbreak{}band15;
\item the error-corrected estimates of Section 3.9.
\end{itemize}
Added after the full-frame estimates were seen (D19): rq2\_\allowbreak{}net\_\allowbreak{}band and rq2\_\allowbreak{}net\_\allowbreak{}band15, the RQ2 net-of-chance contrasts with the random-band control and on repositories matched within 15 per cent of size; the known-groups checks with repository fixed effects (kg\_\allowbreak{}agent\_\allowbreak{}fe, kg\_\allowbreak{}human\_\allowbreak{}fe); and precision and recall by size third.

The sensitivities added after the full-frame estimates were seen (D22), grouped by what they vary:

\begin{itemize}
\item \textbf{Terms the RQ2 model carries.} The pre-specified RQ3 model with fixed effects for the agent group (the human arm as one group), for the month, and for both (rq3\_\allowbreak{}mk\_\allowbreak{}group, rq3\_\allowbreak{}mk\_\allowbreak{}month, rq3\_\allowbreak{}mk\_\allowbreak{}group\_\allowbreak{}month).
\item \textbf{Same repository and month.} RQ3 within the same repository and month, with one fixed effect per repository-month cell and size kept as a term, alone, with PR size and with PR size and the agent group (rq3\_\allowbreak{}um, rq3\_\allowbreak{}um\_\allowbreak{}lines, rq3\_\allowbreak{}um\_\allowbreak{}grp), which compares agent and human pull requests reviewed in the same repository in the same month, and the last without the delegators' comments (rq3\_\allowbreak{}um\_\allowbreak{}grp\_\allowbreak{}nodeleg).
\item \textbf{Period.} The pre-specified RQ3 model on pull requests created before and from 2025-11-01 (rq3\_\allowbreak{}mk\_\allowbreak{}pre, rq3\_\allowbreak{}mk\_\allowbreak{}post), and RQ3 with a three-way term of arm, size and the period from 2025-11-01, which tests whether the arm difference changes between the periods (rq3\_\allowbreak{}mk\_\allowbreak{}period).
\item \textbf{Pull-request size and coverage.} The pre-specified RQ3 model with PR size, or with the pull request's changed files as a share of the tree (pr\_\allowbreak{}cover, log10 of changed files plus one over code files), added as a term (rq3\_\allowbreak{}mk\_\allowbreak{}pr\_\allowbreak{}lines, rq3\_\allowbreak{}mk\_\allowbreak{}cover); RQ3 with repository fixed effects and month effects, or with PR size, each alone (rq3\_\allowbreak{}fe\_\allowbreak{}month, rq3\_\allowbreak{}fe\_\allowbreak{}pr\_\allowbreak{}lines); and RQ2 with pr\_\allowbreak{}cover (rq2\_\allowbreak{}cover).
\item \textbf{Bots without the [bot] suffix.} The pre-specified RQ2 and RQ3 models without comments from accounts that are bots or apps but whose login does not end in [bot], matched by name (rq2\_\allowbreak{}nobots, rq3\_\allowbreak{}mk\_\allowbreak{}nobots).
\item \textbf{Delegators.} RQ2 and RQ3 without the delegators' comments (rq2\_\allowbreak{}nodeleg, rq3\_\allowbreak{}mk\_\allowbreak{}nodeleg, and rq3\_\allowbreak{}mk\_\allowbreak{}nodeleg\_\allowbreak{}gm with the agent group and the month held fixed), and the known-groups check of the agent arm without them (kg\_\allowbreak{}agent\_\allowbreak{}nodeleg). The same RQ2 and RQ3 models without only the delegators found as an assignee or in a ``Requested by'' line (rq2\_\allowbreak{}nodeleg\_\allowbreak{}strict, rq3\_\allowbreak{}mk\_\allowbreak{}nodeleg\_\allowbreak{}strict), a narrower rule that leaves out those who only requested or were requested for review.
\item \textbf{Who wrote the human pull requests.} RQ3 with the human arm restricted to pull requests whose authors are owners, members or collaborators, or to the others (rq3\_\allowbreak{}mk\_\allowbreak{}vs\_\allowbreak{}insiders, rq3\_\allowbreak{}mk\_\allowbreak{}vs\_\allowbreak{}outsiders).
\item \textbf{Pull request as the unit.} The pre-specified RQ2 and RQ3 models with one row per pull request, the outcome being whether any of its comments names an outside item (rq2\_\allowbreak{}pr, rq3\_\allowbreak{}mk\_\allowbreak{}pr).
\item \textbf{Repository traits.} RQ2 with the repository's stars, its age at the pull request and the number of human pull-request authors in the window added as terms (rq2\_\allowbreak{}community), from the GitHub API and the collected pull requests; RQ3 with these traits and their interactions with the arm (rq3\_\allowbreak{}mk\_\allowbreak{}traits); and RQ3 with a CODEOWNERS file at the base commit and its interaction with the arm (rq3\_\allowbreak{}mk\_\allowbreak{}codeowners). RQ2 with fixed effects for the repository's primary language (rq2\_\allowbreak{}lang), with the number of distinct reviewers in the repository (rq2\_\allowbreak{}reviewers), and with the number of comments on the pull request (rq2\_\allowbreak{}prcomments), each added beside the RQ2 terms.
\item \textbf{Addressed comments and weighting.} RQ1 and RQ2 with the comments addressed to an agent restored (rq1\_\allowbreak{}with\_\allowbreak{}addressed, rq2\_\allowbreak{}with\_\allowbreak{}addressed), and RQ1 weighted by repository, the mean of each repository's share (rq1\_\allowbreak{}repo).
\item \textbf{Agent groups.} RQ2 within each of the three largest agent groups (rq2\_\allowbreak{}copilot, rq2\_\allowbreak{}claude, rq2\_\allowbreak{}codex), RQ3 with the agent arm restricted to each of them (rq3\_\allowbreak{}mk\_\allowbreak{}copilot, rq3\_\allowbreak{}mk\_\allowbreak{}claude, rq3\_\allowbreak{}mk\_\allowbreak{}codex), and RQ4 with agent-group effects (rq4\_\allowbreak{}grp).
\item \textbf{Context files.} RQ4 with Copilot's context file also counting AGENTS.md from 2025-08-28, the date GitHub's public changelog gives for Copilot's coding agent beginning to read it (rq4\_\allowbreak{}v2).
\item \textbf{The gradient split in two.} Whether a comment names any resolvable item of the repository, inside the change or outside it (any\_\allowbreak{}ref, the instrument run with the change ignored), and, among comments that do, whether the item lies outside the change (rq2\_\allowbreak{}any, rq2\_\allowbreak{}outside\_\allowbreak{}given\_\allowbreak{}any).
\item \textbf{Review coverage.} The share of pull requests with any inline review, by arm and third of repository size, from the collected pull requests.
\item \textbf{Functional form.} RQ2 with a term for the square of lsize (rq2\_\allowbreak{}quad), and the human arm's model in the comparison set with the same term (rq2\_\allowbreak{}set\_\allowbreak{}human\_\allowbreak{}quad); RQ2 on repositories below 1,500 code files at the base commit and on those at or above it (rq2\_\allowbreak{}below1500, rq2\_\allowbreak{}above1500), a split chosen after the size quintiles had been seen; and RQ2 on repositories with at least 300 code files (rq2\_\allowbreak{}ge300). For the arm comparison, the pre-specified model, the model with agent and month effects, and the repository-month model with and without the delegators' comments, each with the square of centred size shared by both arms (q\_\allowbreak{}reg\_\allowbreak{}shared, q\_\allowbreak{}mk\_\allowbreak{}group\_\allowbreak{}month\_\allowbreak{}shared, q\_\allowbreak{}um\_\allowbreak{}grp\_\allowbreak{}shared, q\_\allowbreak{}um\_\allowbreak{}grp\_\allowbreak{}nodeleg\_\allowbreak{}shared) or separate for each arm (q\_\allowbreak{}reg\_\allowbreak{}arm, q\_\allowbreak{}mk\_\allowbreak{}group\_\allowbreak{}month\_\allowbreak{}arm, q\_\allowbreak{}um\_\allowbreak{}grp\_\allowbreak{}arm, q\_\allowbreak{}um\_\allowbreak{}grp\_\allowbreak{}nodeleg\_\allowbreak{}arm); with a square per arm, the ratio is read at the mean log size and at 300, 1,000 and 3,000 code files (Table B3). The same four specifications with a linear spline in log size per arm, knots at 300 and 3,000 code files, give each arm's slope in small, mid-sized and large repositories (q\_\allowbreak{}reg\_\allowbreak{}pw, q\_\allowbreak{}mk\_\allowbreak{}group\_\allowbreak{}month\_\allowbreak{}pw, q\_\allowbreak{}um\_\allowbreak{}grp\_\allowbreak{}pw, q\_\allowbreak{}um\_\allowbreak{}grp\_\allowbreak{}nodeleg\_\allowbreak{}pw; Table B4).
\item \textbf{Comments formatted as AI review output.} The pre-specified RQ2 and RQ3 models without comments whose text is formatted as AI review tools format their output (rq2\_\allowbreak{}noai, rq3\_\allowbreak{}mk\_\allowbreak{}noai).
\item \textbf{Pull request as the unit, counted.} The pre-specified RQ3 model with one row per pull request, the count of outside references as the outcome and the pull request's comments as exposure (rq3\_\allowbreak{}mk\_\allowbreak{}prcount).
\item \textbf{Human authors by standing.} The pre-specified RQ3 model with the human gradient allowed to differ between insider and non-member authors (rq3\_\allowbreak{}mk\_\allowbreak{}authorsplit).
\item \textbf{Fixed-effects sample.} The pre-specified RQ3 model fitted on the repositories the fixed-effects logit keeps (rq3\_\allowbreak{}registered\_\allowbreak{}fesample).
\item \textbf{Each arm's own gradient.} In the comparison set, a logistic regression of ref\_\allowbreak{}out on lsize and reviewer insider status with month effects for each arm (rq2\_\allowbreak{}set\_\allowbreak{}human, rq2\_\allowbreak{}set\_\allowbreak{}agent), the same model without other terms (rq2\_\allowbreak{}set\_\allowbreak{}human\_\allowbreak{}raw, rq2\_\allowbreak{}set\_\allowbreak{}agent\_\allowbreak{}raw), and the net-of-chance companions on the linear probability scale, without other terms (set\_\allowbreak{}human\_\allowbreak{}net, set\_\allowbreak{}agent\_\allowbreak{}net) and with month effects and reviewer insider status (set\_\allowbreak{}human\_\allowbreak{}net\_\allowbreak{}adj, set\_\allowbreak{}agent\_\allowbreak{}net\_\allowbreak{}adj); the human gradient is tested against RQ2's bound of 1.15. A Mundlak form of each arm's model adds the repository mean of size (rq2\_\allowbreak{}set\_\allowbreak{}human\_\allowbreak{}mk, rq2\_\allowbreak{}set\_\allowbreak{}agent\_\allowbreak{}mk), whose repository-mean term is the difference between the between-repository and the within-repository slopes; the within slope plus that term gives the between-repository slope. The RQ2 net-of-chance contrast is also fitted with the RQ2 model's adjustments, agent and month effects and reviewer insider status (rq2\_\allowbreak{}net\_\allowbreak{}adj). The RQ3 net-of-chance companion is also fitted with month effects (rq3\_\allowbreak{}net\_\allowbreak{}month).
\item \textbf{Convergent check.} The check the plan named (D16): for every flagged comment, whether the files it names, outside the change when it was written, are among the files the pull request changes by its end. We compare them with two baselines drawn at the base commit outside the change and outside the named files (seed 20261010): as many files drawn at random from the whole tree, and, for each named file, one file drawn at random from the same directory, since files next to a change are changed together more often whatever a reviewer writes. A named file whose directory holds no other eligible file has no same-directory partner. Intervals come from a cluster bootstrap over repositories.
\end{itemize}
{\footnotesize
\setlength{\tabcolsep}{4pt}
\begin{tabularx}{\linewidth}{XlX}
\caption{Every estimate and its role.}\\
\toprule
\textbf{Estimate} & \textbf{Question} & \textbf{Role} \\
\midrule\endfirsthead
\toprule \textbf{Estimate} & \textbf{Question} & \textbf{Role} \\ \midrule\endhead
rq1, rq2\_\allowbreak{}registered, rq3\_\allowbreak{}registered & RQ1, RQ2, RQ3 & pre-specified; decision tests for RQ2 and RQ3 \\
rq2\_\allowbreak{}net, rq3\_\allowbreak{}net & RQ2, RQ3 & net-of-chance companions (next-partner control) \\
rq2\_\allowbreak{}net\_\allowbreak{}cf, rq3\_\allowbreak{}net\_\allowbreak{}cf & RQ2, RQ3 & sensitivity (next partner re-paired on code files, D20) \\
rq2, rq3, rq3\_\allowbreak{}frozen, rq3\_\allowbreak{}with\_\allowbreak{}addressed, rq3\_\allowbreak{}size\_\allowbreak{}time, rq3\_\allowbreak{}no\_\allowbreak{}truncated, rq3\_\allowbreak{}aidev\_\allowbreak{}window & RQ2, RQ3 & sensitivity (Section 3.7) \\
rq3\_\allowbreak{}net\_\allowbreak{}band, rq3\_\allowbreak{}net\_\allowbreak{}band15 & RQ3 & sensitivity (random-band control) \\
rq2\_\allowbreak{}net\_\allowbreak{}band, rq2\_\allowbreak{}net\_\allowbreak{}band15 & RQ2 & sensitivity (random-band control, D19) \\
DSL models for RQ2 and RQ3 & RQ2, RQ3 & sensitivity (Section 3.9) \\
kg\_\allowbreak{}agent, kg\_\allowbreak{}human & validity & known-groups check \\
kg\_\allowbreak{}agent\_\allowbreak{}fe, kg\_\allowbreak{}human\_\allowbreak{}fe & validity & known-groups check with repository fixed effects (D19) \\
kg\_\allowbreak{}agent\_\allowbreak{}nodeleg & validity & known-groups check without delegators (D22) \\
convergent check & validity & convergent validity check: named files are later changed more often than comparable files (D16, D22) \\
all rq3\_\allowbreak{}mk\_\allowbreak{} and rq3\_\allowbreak{}um models, rq3\_\allowbreak{}fe\_\allowbreak{}month, rq3\_\allowbreak{}fe\_\allowbreak{}pr\_\allowbreak{}lines, rq1\_\allowbreak{}with\_\allowbreak{}addressed, rq2\_\allowbreak{}with\_\allowbreak{}addressed, rq2\_\allowbreak{}cover, rq2\_\allowbreak{}nobots, rq2\_\allowbreak{}pr, rq2\_\allowbreak{}community, rq2\_\allowbreak{}nodeleg & RQ1, RQ2, RQ3 & sensitivity (D22), decision rule reported for description only, not part of the confirmatory inference \\
rq2\_\allowbreak{}set\_\allowbreak{}human, rq2\_\allowbreak{}set\_\allowbreak{}agent, rq2\_\allowbreak{}set\_\allowbreak{}human\_\allowbreak{}raw, rq2\_\allowbreak{}set\_\allowbreak{}agent\_\allowbreak{}raw, set\_\allowbreak{}human\_\allowbreak{}net, set\_\allowbreak{}agent\_\allowbreak{}net, set\_\allowbreak{}human\_\allowbreak{}net\_\allowbreak{}adj, set\_\allowbreak{}agent\_\allowbreak{}net\_\allowbreak{}adj, rq2\_\allowbreak{}set\_\allowbreak{}human\_\allowbreak{}mk, rq2\_\allowbreak{}set\_\allowbreak{}agent\_\allowbreak{}mk & RQ3 & each arm's gradient, raw and adjusted, net of chance, and its within- and between-repository slopes (D22) \\
rq1\_\allowbreak{}repo & RQ1 & sensitivity (D22): the share weighted by repository \\
rq2\_\allowbreak{}copilot, rq2\_\allowbreak{}claude, rq2\_\allowbreak{}codex, rq3\_\allowbreak{}mk\_\allowbreak{}copilot, rq3\_\allowbreak{}mk\_\allowbreak{}claude, rq3\_\allowbreak{}mk\_\allowbreak{}codex & RQ2, RQ3 & sensitivity (D22): each of the three largest agent groups \\
rq2\_\allowbreak{}any, rq2\_\allowbreak{}outside\_\allowbreak{}given\_\allowbreak{}any & RQ2 & sensitivity (D22): the gradient split into naming any item and the item lying outside the change \\
rq2\_\allowbreak{}lang, rq2\_\allowbreak{}reviewers, rq2\_\allowbreak{}prcomments, rq2\_\allowbreak{}net\_\allowbreak{}adj & RQ2 & sensitivity (D22): language, reviewers per repository and comments per pull request; the net-of-chance contrast with the RQ2 adjustments \\
rq2\_\allowbreak{}nodeleg\_\allowbreak{}strict, rq3\_\allowbreak{}mk\_\allowbreak{}nodeleg\_\allowbreak{}strict & RQ2, RQ3 & sensitivity (D22): delegators identified by assignees and ``Requested by'' only \\
rq2\_\allowbreak{}quad, rq2\_\allowbreak{}set\_\allowbreak{}human\_\allowbreak{}quad, rq2\_\allowbreak{}below1500, rq2\_\allowbreak{}above1500, rq2\_\allowbreak{}ge300 & RQ2, RQ3 & sensitivity (D22): functional form of the size gradient and the trimmed range \\
rq2\_\allowbreak{}noai, rq3\_\allowbreak{}mk\_\allowbreak{}noai & RQ2, RQ3 & sensitivity (D22): comments formatted as AI review output left out \\
rq3\_\allowbreak{}mk\_\allowbreak{}prcount, rq3\_\allowbreak{}mk\_\allowbreak{}authorsplit, rq3\_\allowbreak{}registered\_\allowbreak{}fesample & RQ3 & sensitivity (D22): pull-request count model; human gradient by author standing; pre-specified model on the fixed-effects sample \\
review coverage, descriptive tables & RQ2, RQ3 & description (D22): share of pull requests reviewed, composition of the arms, pull-request size and comments per pull request by arm and size third \\
rq4, rq4\_\allowbreak{}v2, rq4\_\allowbreak{}grp & RQ4 & exploratory; no confirmatory inference \\
rq3\_\allowbreak{}net\_\allowbreak{}month & RQ3 & sensitivity (D22): the net-of-chance companion with month effects \\
q\_\allowbreak{}reg\_\allowbreak{}shared, q\_\allowbreak{}reg\_\allowbreak{}arm, q\_\allowbreak{}mk\_\allowbreak{}group\_\allowbreak{}month\_\allowbreak{}shared, q\_\allowbreak{}mk\_\allowbreak{}group\_\allowbreak{}month\_\allowbreak{}arm, q\_\allowbreak{}um\_\allowbreak{}grp\_\allowbreak{}shared, q\_\allowbreak{}um\_\allowbreak{}grp\_\allowbreak{}arm, q\_\allowbreak{}um\_\allowbreak{}grp\_\allowbreak{}nodeleg\_\allowbreak{}shared, q\_\allowbreak{}um\_\allowbreak{}grp\_\allowbreak{}nodeleg\_\allowbreak{}arm, q\_\allowbreak{}reg\_\allowbreak{}pw, q\_\allowbreak{}mk\_\allowbreak{}group\_\allowbreak{}month\_\allowbreak{}pw, q\_\allowbreak{}um\_\allowbreak{}grp\_\allowbreak{}pw, q\_\allowbreak{}um\_\allowbreak{}grp\_\allowbreak{}nodeleg\_\allowbreak{}pw & RQ3 & sensitivity (D22): the arm comparison with a curved size gradient (Tables B3 and B4) \\
\bottomrule
\end{tabularx}}
{\footnotesize
\setlength{\tabcolsep}{4pt}
\begin{tabularx}{\linewidth}{Xl>{\raggedleft\arraybackslash}l>{\raggedleft\arraybackslash}l>{\raggedleft\arraybackslash}l>{\raggedleft\arraybackslash}lX}
\caption{The arm comparison under every specification fitted on the full comparison set (Appendix F.4 notes the pre-specified model on the fixed-effects sample): the ratio of the agent to the human size odds ratio (for the count model, of the rate ratios), its 95 per cent interval, the 90 per cent interval that decides the equivalence test, and the pre-specified decision rule applied to it (equivalence bound 1.25; Holm adjustment for the pre-specified model only). Status says whether a row was pre-specified, a planned sensitivity (D11) or added after the results were seen (D22). Rows with a square of size per arm give the ratio at the mean log size over the analysis frame, about 845 code files (Section 3.4). For these rows, the ratio is the exponentiated difference between the arms' local log-odds slopes in log size at that size, not a ratio of odds ratios for a finite tenfold change. Rows that do not name another base model (the same repository and month, repository fixed effects or the earlier round's formula) modify the pre-specified model. Upper ends of the 90 per cent interval, and a lower end near the bound, are given to four decimals. † marks a TOST p value between 0.045 and 0.055, on the decision boundary, where a small change in the data would change the verdict.}\\
\toprule
\textbf{Specification} & \textbf{Status} & \textbf{Ratio} & \textbf{95 per cent interval} & \textbf{90 per cent interval} & \textbf{TOST p} & \textbf{Verdict} \\
\midrule\endfirsthead
\toprule \textbf{Specification} & \textbf{Status} & \textbf{Ratio} & \textbf{95 per cent interval} & \textbf{90 per cent interval} & \textbf{TOST p} & \textbf{Verdict} \\ \midrule\endhead
pre-specified (Mundlak) † & pre-specified & 1.15 & 1.04 to 1.27 & 1.06 to 1.2490 & 0.0484 & different from zero, and smaller than the smallest effect of interest \\
same repository and month & added after results (D22) & 1.06 & 0.92 to 1.22 & 0.94 to 1.1891 & 0.0098 & equivalent to zero within the smallest effect of interest \\
same repository and month + PR size & added after results (D22) & 1.06 & 0.92 to 1.22 & 0.94 to 1.1942 & 0.0116 & equivalent to zero within the smallest effect of interest \\
same repository and month + PR size + agent group & added after results (D22) & 1.06 & 0.92 to 1.22 & 0.94 to 1.1936 & 0.0112 & equivalent to zero within the smallest effect of interest \\
same repository and month + PR size + agent group, delegators' comments left out & added after results (D22) & 1.11 & 0.94 to 1.31 & 0.96 to 1.2772 & 0.0819 & inconclusive: neither different from zero nor within the smallest effect of interest \\
pre-specified + agent group & added after results (D22) & 1.08 & 0.98 to 1.19 & 0.99 to 1.1731 & 0.0018 & equivalent to zero within the smallest effect of interest \\
pre-specified + month & added after results (D22) & 1.02 & 0.93 to 1.11 & 0.95 to 1.0988 & < 0.0001 & equivalent to zero within the smallest effect of interest \\
pre-specified + agent group and month & added after results (D22) & 0.99 & 0.91 to 1.09 & 0.92 to 1.0707 & < 0.0001 & equivalent to zero within the smallest effect of interest \\
before November 2025 & added after results (D22) & 1.03 & 0.91 to 1.17 & 0.93 to 1.1477 & 0.0015 & equivalent to zero within the smallest effect of interest \\
from November 2025 & added after results (D22) & 1.06 & 0.95 to 1.19 & 0.97 to 1.1701 & 0.0027 & equivalent to zero within the smallest effect of interest \\
one row per pull request: any outside reference & added after results (D22) & 1.05 & 0.96 to 1.13 & 0.98 to 1.1190 & < 0.0001 & equivalent to zero within the smallest effect of interest \\
one row per pull request: references per comment (count, comments as exposure) & added after results (D22) & 1.24 & 1.09 to 1.40 & 1.12 to 1.3722 & 0.4379 & different from zero \\
pre-specified + PR size † & added after results (D22) & 1.15 & 1.04 to 1.27 & 1.06 to 1.2499 & 0.0499 & different from zero, and smaller than the smallest effect of interest \\
pre-specified + CODEOWNERS and its interaction with the arm & added after results (D22) & 1.23 & 1.06 to 1.44 & 1.08 to 1.4028 & 0.4312 & different from zero \\
pre-specified + PR's share of the tree † & added after results (D22) & 1.15 & 1.04 to 1.27 & 1.06 to 1.2484 & 0.0474 & different from zero, and smaller than the smallest effect of interest \\
bot-like accounts left out † & added after results (D22) & 1.15 & 1.05 to 1.27 & 1.07 to 1.2503 & 0.0505 & different from zero \\
comments formatted as AI review output left out & added after results (D22) & 1.17 & 1.06 to 1.28 & 1.08 to 1.2640 & 0.0786 & different from zero \\
delegators' comments left out & added after results (D22) & 1.17 & 1.04 to 1.31 & 1.06 to 1.2843 & 0.1179 & different from zero \\
delegators' comments left out, assignees and ``Requested by'' only & added after results (D22) & 1.17 & 1.05 to 1.30 & 1.07 to 1.2753 & 0.1008 & different from zero \\
delegators' comments left out, + agent group and month & added after results (D22) & 1.02 & 0.93 to 1.13 & 0.94 to 1.1108 & < 0.0001 & equivalent to zero within the smallest effect of interest \\
pre-specified + repository traits and their interactions with the arm & added after results (D22) & 1.14 & 1.00 to 1.31 & 1.02 to 1.2779 & 0.0924 & inconclusive: neither different from zero nor within the smallest effect of interest \\
agent arm: Copilot only & added after results (D22) & 1.09 & 0.96 to 1.25 & 0.98 to 1.2211 & 0.0233 & equivalent to zero within the smallest effect of interest \\
agent arm: Claude Code only & added after results (D22) & 1.08 & 0.93 to 1.24 & 0.96 to 1.2154 & 0.0212 & equivalent to zero within the smallest effect of interest \\
agent arm: OpenAI Codex only & added after results (D22) & 1.32 & 1.02 to 1.72 & 1.07 to 1.6463 & 0.6705 & different from zero \\
human arm: insider authors only & added after results (D22) & 1.10 & 0.98 to 1.22 & 1.00 to 1.2027 & 0.0099 & equivalent to zero within the smallest effect of interest \\
human arm: non-member authors only & added after results (D22) & 1.22 & 1.07 to 1.38 & 1.10 to 1.3528 & 0.3405 & different from zero \\
repository fixed effects & planned (D11) & 1.17 & 1.02 to 1.34 & 1.04 to 1.3127 & 0.1731 & different from zero \\
repository fixed effects + month & added after results (D22) & 1.04 & 0.93 to 1.17 & 0.94 to 1.1451 & 0.0009 & equivalent to zero within the smallest effect of interest \\
repository fixed effects + PR size & added after results (D22) & 1.17 & 1.02 to 1.35 & 1.05 to 1.3169 & 0.1843 & different from zero \\
repository fixed effects + PR size + month & planned (D11) & 1.04 & 0.93 to 1.17 & 0.95 to 1.1495 & 0.0011 & equivalent to zero within the smallest effect of interest \\
repository fixed effects, before November 2025 & planned (D11) & 1.03 & 0.88 to 1.22 & 0.90 to 1.1842 & 0.0109 & equivalent to zero within the smallest effect of interest \\
addressed comments restored & planned (D11) & 1.15 & 1.01 to 1.31 & 1.03 to 1.2796 & 0.0988 & different from zero \\
pre-specified + a square of size shared by both arms & added after results (D22) & 1.14 & 1.00 to 1.29 & 1.02 to 1.2634 & 0.0695 & different from zero \\
pre-specified + a square of size per arm (ratio at the mean log size) & added after results (D22) & 1.23 & 1.07 to 1.42 & 1.09 to 1.3919 & 0.4221 & different from zero \\
pre-specified + agent group and month + a square of size shared by both arms & added after results (D22) & 1.00 & 0.89 to 1.11 & 0.91 to 1.0921 & < 0.0001 & equivalent to zero within the smallest effect of interest \\
pre-specified + agent group and month + a square of size per arm (ratio at the mean log size) & added after results (D22) & 1.06 & 0.93 to 1.21 & 0.95 to 1.1808 & 0.0060 & equivalent to zero within the smallest effect of interest \\
same repository and month + PR size + agent group + a square of size shared by both arms & added after results (D22) & 1.06 & 0.92 to 1.22 & 0.94 to 1.1908 & 0.0103 & equivalent to zero within the smallest effect of interest \\
same repository and month + PR size + agent group + a square of size per arm (ratio at the mean log size) & added after results (D22) & 1.22 & 1.04 to 1.43 & 1.07 to 1.3947 & 0.3817 & different from zero \\
same repository and month + PR size + agent group + a square of size shared by both arms, delegators' comments left out & added after results (D22) & 1.11 & 0.94 to 1.31 & 0.96 to 1.2730 & 0.0760 & inconclusive: neither different from zero nor within the smallest effect of interest \\
same repository and month + PR size + agent group + a square of size per arm, delegators' comments left out & added after results (D22) & 1.30 & 1.07 to 1.58 & 1.11 to 1.5344 & 0.6616 & different from zero \\
earlier round's formula, no size main effect & planned (D11) & 1.43 & 1.22 to 1.67 & 1.2509 to 1.6319 & 0.9510 & different from zero \\
\bottomrule
\end{tabularx}}
\needspace{14\baselineskip}
{\footnotesize
\setlength{\tabcolsep}{4pt}
\begin{tabularx}{\linewidth}{XXXll}
\caption{The arm comparison with a square of size per arm, read at three sizes: the ratio of the agent to the human size odds ratio at 300, 1,000 and 3,000 code files at the base commit, with its 95 per cent interval and the upper end of its 90 per cent interval (against the bound of 1.25), and the difference between the arms in the squared term on the log-odds scale. All rows were added after the results (D22).}\\
\toprule
\textbf{Specification} & \textbf{Difference in the squared term} & \textbf{At 300 code files} & \textbf{At 1,000 code files} & \textbf{At 3,000 code files} \\
\midrule\endfirsthead
\toprule \textbf{Specification} & \textbf{Difference in the squared term} & \textbf{At 300 code files} & \textbf{At 1,000 code files} & \textbf{At 3,000 code files} \\ \midrule\endhead
pre-specified (Mundlak) & -0.10 (-0.23 to 0.02) & 1.35 (1.08 to 1.68; 90 per cent upper 1.6260) & 1.21 (1.06 to 1.39; 1.3630) & 1.10 (0.95 to 1.28; 1.2487) \\
pre-specified + agent group and month & -0.09 (-0.20 to 0.02) & 1.15 (0.95 to 1.40; 1.3596) & 1.05 (0.93 to 1.18; 1.1580) & 0.96 (0.84 to 1.08; 1.0632) \\
same repository and month + PR size + agent group & -0.20 (-0.32 to -0.07) & 1.45 (1.14 to 1.85; 1.7784) & 1.19 (1.02 to 1.38; 1.3455) & 0.98 (0.85 to 1.14; 1.1139) \\
as above, delegators' comments left out & -0.21 (-0.36 to -0.05) & 1.57 (1.16 to 2.11; 2.0148) & 1.26 (1.05 to 1.52; 1.4744) & 1.04 (0.87 to 1.24; 1.2052) \\
\bottomrule
\end{tabularx}}
\needspace{13\baselineskip}
{\small
\setlength{\tabcolsep}{4pt}
\begin{tabularx}{\linewidth}{Xlll}
\caption{The arm comparison with a linear spline in log size per arm, knots at 300 and 3,000 code files: the ratio of the agent to the human size odds ratio within small (below 300 code files), mid-sized (300 to 3,000) and large (above 3,000) repositories, with its 95 per cent interval. Under the pre-specified decision rule (equivalence bound 1.25; applied without Holm adjustment, as for every sensitivity), the small-repository ratio within the same repository and month is ``different from zero'' on its unadjusted p (0.019), but not after Holm adjustment over the three size classes (0.057); every other verdict is inconclusive. All rows were added after the results (D22).}\\
\toprule
\textbf{Specification} & \textbf{Small} & \textbf{Mid-sized} & \textbf{Large} \\
\midrule\endfirsthead
\toprule \textbf{Specification} & \textbf{Small} & \textbf{Mid-sized} & \textbf{Large} \\ \midrule\endhead
pre-specified (Mundlak) & 1.33 (0.93 to 1.90) & 1.21 (0.93 to 1.57) & 1.03 (0.73 to 1.46) \\
pre-specified + agent group and month & 1.36 (0.92 to 2.01) & 0.99 (0.75 to 1.30) & 0.92 (0.68 to 1.24) \\
same repository and month + PR size + agent group & 1.99 (1.12 to 3.53) & 1.15 (0.84 to 1.59) & 0.81 (0.59 to 1.13) \\
as above, delegators' comments left out & 2.03 (0.99 to 4.16) & 1.25 (0.86 to 1.82) & 0.85 (0.58 to 1.24) \\
\bottomrule
\end{tabularx}}
The results below are those of the sensitivities that Section 4 summarises in a sentence.

\textbf{The arm comparison under other specifications (Table B2, Figure 3).} The two headline comparisons estimate related but different quantities. The pre-specified model pools over agents and months, which both vary with repository size. Copilot makes up a smaller share of agent comments in large repositories and OpenAI Codex a larger one, and later months hold more agent comments. The repository-month comparison conditions on both. The excess disappears when the agent group, the month or both are held fixed (1.08, 1.02 and 0.99). It also disappears within the periods before and after November 2025 (1.03 and 1.06), and the arm difference does not change between those periods (three-way odds ratio 1.09, interval 0.86 to 1.38). The result with month effects is consistent with the excess reflecting how the month mix varies with size, though holding the month fixed may also remove part of an agent effect (Appendix F.2). The excess remains under repository fixed effects (1.17), with PR size (1.15) or CODEOWNERS (1.23), and with addressed comments restored (1.15). It also remains with bot-like accounts (1.15), AI-formatted comments (1.17) or the delegators' comments (1.17) left out. When a pull request's outside references are counted with its comments as exposure, the ratio is 1.24 (interval 1.09 to 1.40). In several of these specifications the equivalence test does not rule out a ratio of 1.25. Whether any of a pull request's comments names an outside item gives 1.05, because agent pull requests in larger repositories receive fewer comments. With repository traits and their interactions with the arm added, the ratio is 1.14 (interval 1.00 to 1.31), ``inconclusive: neither different from zero nor within the smallest effect of interest''.

\textbf{The arm comparison with a curved size gradient (Tables B3 and B4).} The headline comparisons fit each arm's gradient as a straight line in log size, while Section 4.2 found the gradient concave. Added after the results (D22), each headline comparison was refitted with the square of size, first shared by both arms and then separate for each (Table B3). With a square of centred log size per arm, the ratio at x code files is exp(b\_\allowbreak{}lin + 2·b\_\allowbreak{}sq·(log10 x − c)). Here b\_\allowbreak{}lin and b\_\allowbreak{}sq are the arm differences in the linear and the squared term, and c is the frame mean of log10 size (about 845 code files). A shared square leaves the ratios almost unchanged: 1.14 in the pre-specified model, 1.00 with agent-group and month effects added, and 1.06 within the same repository and month (1.11 without the delegators' comments). In the pre-specified model, though, the equivalence test no longer passes (p = 0.0695). A square per arm changes the ratios. The pre-specified ratio at the mean log size becomes 1.23 (interval 1.07 to 1.42), and the equivalence test no longer passes (p = 0.4221). Within the same repository and month it becomes 1.22 (interval 1.04 to 1.43). In both, the agents' rise is steeper at 300 and 1,000 code files, and neither detects an excess at 3,000 (Table B3). The two arms' curvatures differ detectably within the same repository and month (the difference in the squared term is -0.20 on the log-odds scale, interval -0.32 to -0.07). They do not in the pre-specified model (-0.10, interval -0.23 to 0.02). With agent-group and month effects added to the pre-specified model, the ratio is 1.06 at the mean log size and 1.15 at 300 code files (interval 0.95 to 1.40).

A quadratic has one curvature term per arm, so it places any arm difference toward one end of the size range by construction. A linear spline in log size, with a separate slope for each arm in small (below 300 code files), mid-sized and large (above 3,000) repositories, is not bound that way (Table B4). With the spline, the ratio in a size class is the exponential of the arm difference in slope within that class. For the pre-specified comparison, none of the three ratios is detectably above 1. Within the same repository and month, only the small-repository ratio is: p = 0.019 unadjusted, and 0.057 after Holm adjustment over the three size classes. Without the delegators' comments, the small-repository ratio is 2.03 (interval 0.99 to 4.16; p = 0.053). The verdicts on the arm comparison therefore depend on the functional form of size as well as on the specification. As a log-linear average, the repository-month comparison is equivalent to 1. With a quadratic per arm, both comparisons find the agents' rise steeper among small and mid-sized repositories, unless agent-group and month effects are added to the pre-specified model. With a slope per size class, an excess is nominally detected only among small repositories, within the same repository and month. Both forms give the same ordering, the ratio falling from small to large repositories; the fall is detectable within the same repository and month under the quadratic (Table B3). The two forms differ in whether the excess in mid-sized repositories is detectable.

\textbf{Each agent group and human authors by standing.} Compared agent by agent with the same human comments, the ratio is 1.09 for Copilot (interval 0.96 to 1.25), 1.08 for Claude Code (interval 0.93 to 1.24) and 1.32 for OpenAI Codex (interval 1.02 to 1.72). Owners, members or collaborators open 38 per cent of the human pull requests. Against their pull requests, the ratio is 1.10 (interval 0.98 to 1.22). Against those by non-member authors, most of whom GitHub labelled as earlier contributors at collection, it is 1.22 (interval 1.07 to 1.38). In a model that lets the human gradient differ by author standing, the human gradient is 1.05 times steeper for insider than for non-member authors (interval 0.93 to 1.19). A shared distance from the project would put the ratio against non-member authors nearer 1, and it is not. The direction runs against that reading, though the difference between the two human gradients is small and uncertain (Appendix F.2).

\textbf{Context files (RQ4).} With agent-group effects, the RQ4 ratio is 0.99 (interval 0.74 to 1.33). Counting AGENTS.md for Copilot from 2025-08-28 gives 1.00 (Appendix F.1). Separately from the gradient, at the mean repository size, comments on pull requests opened with the context file present carry an odds ratio of 1.34 for an outside reference (interval 1.06 to 1.70).

\setcounter{table}{0}\renewcommand{\thetable}{C\arabic{table}}
\setcounter{figure}{0}\renewcommand{\thefigure}{C\arabic{figure}}
\section*{Appendix C Validation sample, judges and readers}
\addcontentsline{toc}{section}{Appendix C Validation sample, judges and readers}
To learn how often the instrument is right and how much it misses, we check it on a stratified sample of comments from both arms. The sample has 12 strata: arm (2) by third of repository size (3) by the instrument's verdict (an outside reference found or not). Each stratum's quota of 50 comments takes first, with certainty, the comments of the earlier round's planned precision check that fall in it (the carried-over comments), and fills the rest by a simple random sample of the stratum's other comments (seed 20261010). A carried-over comment's inclusion probability is therefore 1, and another comment's is the stratum's remaining quota over its remaining comments. Every validation estimate weights each comment by the inverse of its inclusion probability (Horvitz and Thompson, 1952). The sample was drawn on the instrument's earlier verdicts and is kept as drawn (D21).

We collect two kinds of evidence: facts from the GitHub API and judgements by language models. The facts come from a source the instrument does not read. For each flagged comment, up to three matched items are checked: that the item exists at the base commit (contents API), that a matched symbol occurs in the file it was resolved to, and that the item lies outside the change, the files the compare API lists between the base commit and the comment's commit that the pull request itself changes. The judgements come from language models, blind to arm and size: gemma4:12b (Google, run locally) and Claude Sonnet 5.5 (Anthropic) as judges, and qwen3:14b (Alibaba, run locally) as adjudicator where the two judges disagree. For a flagged comment the judges answer whether the comment refers to the matched item and whether it directs the contributor to it. For an unflagged comment they answer whether it names a specific item of the repository outside the commented hunk, and list the names.

Each sampled comment is then scored. A flagged comment is scored correct when one of its items passes every fact and the final judgement says the comment refers to it. An unflagged comment is scored a miss when the final judgement says it names an outside item and one of the listed names resolves in the index to an item at the base commit outside the change. This rule keeps a judge's ``yes'' for a library name such as ``python'' from counting as a miss, but it also makes recall conditional on the index, so we report recall under a second rule as well, which counts every miss the final judgement reports.

From these scores we compute weighted precision (the share of flagged comments that are correct), recall (the share of true references the instrument finds) and false-omission rate (the share of unflagged comments that do name an outside item), overall and per arm. Their intervals come from a cluster bootstrap that resamples repositories (1,000 resamples, seed 1), a pairs cluster bootstrap of the kind discussed for errors correlated within groups (Cameron et al., 2008), here applied to weighted precision and recall rather than to regression coefficients. We also compute precision and recall per arm within each third of repository size, since a difference in recall between arms biases a comparison of odds-ratio gradients only if recall changes with size differently in the two arms, while on the percentage-point scale even a constant difference in recall scales each arm's slope.

The local models decode at temperature 0 with answers constrained to JSON, and Claude Sonnet 5.5 ran with its default sampling settings; the prompt of every judged comment is in the replication package. Agreement between language models and people varies strongly from one annotation task to another, so it should be checked for each construct (Ahmed et al., 2025). People therefore check the model judges' labels, on review sheets that show each comment, its diff hunk and a label, blind to arm and size. Into every ten items we plant one or two further items, flagged comments on whose label every judge agreed, shown with that label reversed on purpose; a reader who catches fewer than 70 per cent of these planted reversals is not used. The readers are software engineers who volunteered, read public review comments and were not paid; no personal data beyond each reader's catch rate of the planted reversals, stored under a code, were collected, and that rate decided only whether their labels were used.

\begin{itemize}
\item On the first set of sheets, four readers, the author and three of the engineers named in the acknowledgements, checked the final model labels of items in a random order stratified by the instrument's verdict, from the start of the list.
\item A second set, read by another of the engineers named in the acknowledgements, holds the misses the judges reported but the index could not confirm and a random sample of items not yet checked; it is not stratified. It shows for each of its items the model's label or its reverse at random, so the reader must judge every item; the planted items, always shown reversed, are added as on the first set, and only they count as planted reversals.
\end{itemize}
Where people answered, their label is the reference and takes precedence over the model judgements. For a miss, the person decides whether the comment names an item of the project, and the index whether that item lies outside the change, which the sheet does not show. Every validation estimate records whether it rests on people's labels or on the model reference.

\textbf{Results by size third.} Counting also the judged miss that the index does not confirm, the false-omission rate is 4.7 per cent in the agent arm and 6.3 per cent in the human arm. Across the thirds of repository size, the sample holds 1, 5 and 3 misses in the agent arm and 3, 4 and 5 in the human arm. The counts are unweighted. Weighted by inverse inclusion probabilities, a miss stands for many unsampled comments, so the weighted counts would describe the population, not how many misses the estimates rest on. Recall by third, from the smallest to the largest, is 52.6 (interval 23.2 to 100.0), 36.2 (17.1 to 71.7) and 74.7 (40.0 to 100.0) per cent in the agent arm. In the human arm it is 44.9 (18.7 to 100.0), 51.9 (20.8 to 100.0) and 25.9 (13.3 to 58.5) per cent. These intervals are too wide to show a trend. Precision falls with size in the human arm (84.2, 78.1 and 76.5 per cent) and is flat in the agent arm (82.5, 83.7 and 80.9 per cent). Without the comments carried over from the earlier sample, precision is 79.7 per cent, recall 43.5 per cent and the corrected odds ratios 1.22 and 0.80.

\textbf{People's checks and agreement.} On two sets of review sheets, people made 190 checks of 186 of the validation's labels, some labels being checked on both sets. Between 11 and 20 labels were checked in each of the 12 strata, and people agreed with the final model label on 95 per cent. Both sets carried labels reversed on purpose, to test whether readers judged each item. On the first set, which showed the model's label, 4 readers caught 46 of 52 reversed labels. The second set showed for each item the model's label or its reverse at random, so that the reader had to judge every item, and carried planted reversals as the first did. There, one reader caught 14 of 15 planted reversals and agreed with the model on 92 per cent of 91 items. The two model judges agreed on 90.3 per cent of the comments both judged (Cohen's kappa 0.78; Cohen, 1960). On the first set, all four readers gave the same answer on 93 per cent of the items they all read. Fleiss' kappa (Fleiss, 1971) for the binary label (reference or not), four raters per item, is 0.93. Both kappas measure agreement beyond what chance alone would give.

\textbf{Convergent check.} Of the files named outside the change, 10.9 per cent are among the files the pull request changes by its end: 10.5 per cent on agent and 11.2 per cent on human pull requests. Of the 22,655 named files that have another file in their directory, 10.8 per cent are, against 0.9 per cent of a file drawn from the same directory for each. Of files drawn from anywhere in the tree, 0.2 per cent are.

\textbf{A validation that can see size.} A human-labelled reference stratified by arm and size third could show whether recall changes with size. At each cell's observed false-omission rate, holding about 30 misses in each cell would take about 1,500, 294 and 989 unflagged agent-arm comments. In the human arm it would take 1,229, 690 and 282, in the smallest, middle and largest thirds of size. How precise an error correction is depends mainly on the validation sample and the accuracy of the measure it corrects, not on the number of comments analysed (Egami et al., 2023).

\setcounter{table}{0}\renewcommand{\thetable}{D\arabic{table}}
\setcounter{figure}{0}\renewcommand{\thefigure}{D\arabic{figure}}
\section*{Appendix D Smallest effects of interest}
\addcontentsline{toc}{section}{Appendix D Smallest effects of interest}
The plan fixed a smallest effect of interest (SESOI) for each test, the smallest effect that would matter. For RQ2 it is an odds ratio of 1.15 per tenfold increase, below which the share would change by under one percentage point. For RQ3 it is a ratio of odds ratios of 1.25: the agent size odds ratio 1.25 times the human one, or less than 1/\allowbreak{}1.25 of it. We set the RQ3 bound in the plan at about the size of the whole agent gradient that the earlier round of our study had estimated, an odds ratio of 1.23 measured on a different, model-labelled outcome, so that a difference between the arms counts only when it is as large as a whole gradient. On the percentage-point scale, at the human share of the comparison set, a ratio of 1.25 corresponds to about one more outside reference per hundred comments per tenfold increase in size. That is the same scale as RQ2's bound, below which the share would change by under one percentage point, so both bounds treat a change of about one point per tenfold increase as the smallest that matters. With the agent gradient at its RQ2 value, a ratio at the bound would leave a human odds ratio near 1.08.

\setcounter{table}{0}\renewcommand{\thetable}{E\arabic{table}}
\setcounter{figure}{0}\renewcommand{\thefigure}{E\arabic{figure}}
\section*{Appendix E Each arm's gradient in detail}
\addcontentsline{toc}{section}{Appendix E Each arm's gradient in detail}
In the comparison set, 6.5 per cent of comments on agent and 4.3 per cent on human pull requests name an outside item. Without other terms, the size odds ratio is 1.60 for agent and 1.50 for human pull requests; with reviewer insider status and month effects, it is 1.40 and 1.41. The human gradient is not negligible by RQ2's bound: the equivalence test against 1.15 gives p = 1.000, ``different from zero''.

On the percentage-point scale, net of chance, the agent share rises by 2.03 and the human share by 1.30 points per tenfold increase without other terms, and by 1.27 (interval 0.59 to 1.96) and 1.04 (interval 0.82 to 1.26) with insider status and month effects, so most of the raw gap comes from the mix of months and reviewer standing. The pre-specified net-of-chance companion puts the agent excess at 0.91 points (interval 0.40 to 1.42). Holding the month fixed narrows it to 0.45 points (interval -0.002 to 0.90, just including zero), as on the odds-ratio scale. On this scale, then, the excess also depends on whether the month is held fixed. The instrument's lower recall on human pull requests shrinks the human slope on this scale, so the two arms' percentage-point slopes cannot be compared directly. As an illustration only, rescaled by each arm's precision over recall, the unadjusted rises become 2.80 and 2.63, and the adjusted ones 1.75 and 2.12. The rescaling assumes that precision and recall do not change with size. It carries no interval, although human recall is itself uncertain (its interval runs from 23.3 to 63.1 per cent). It also applies precision to slopes already net of chance, which discounts chance matches twice (Appendix F.1).

\textbf{Agent groups.} In the three largest agent groups, the RQ2 size odds ratio is 1.50 for Claude Code (interval 1.19 to 1.88) and 1.76 for OpenAI Codex (interval 1.17 to 2.65). For Copilot it is 1.12 (interval 0.91 to 1.36). Most comments on Copilot pull requests come from their delegators under the broad rule, which may change what Copilot's reviewers name (Appendix F.1).

\textbf{The gradient split in two.} A rise can come from reviewers naming the repository more often, or from what they name lying outside the change more often. For the first, the odds that a comment names any item the instrument can find in the repository, inside the change or outside it, rise by an odds ratio of 1.28 per tenfold increase (interval 1.10 to 1.50). For the second, among the comments that name such an item, the odds that it lies outside the change rise by 1.38 (interval 1.19 to 1.60).

In a Mundlak form of the per-arm models, the within-repository size odds ratio, identified from how a repository's size changes between base commits and so confounded with everything else that changes over the same months (Appendix F.2), is 3.73 (interval 1.53 to 9.09) for agent and 1.83 (interval 1.30 to 2.57) for human pull requests. The between-repository odds ratio, the within slope plus the repository-mean term, is 1.39 (interval 1.19 to 1.62) and 1.41 (interval 1.31 to 1.50). These within-repository odds ratios are much smaller than those of the pre-specified model in Appendix F.2 (9.26 and 8.04): the per-arm models carry month effects, which absorb much of what changes within a repository over the same months, while the pre-specified model carries none.

The model's predicted shares put these gradients in everyday terms. For a repository at the first, fifth and ninth decile of size, the pre-specified model, which is log-linear in size, predicts that 3.4, 4.5 and 5.8 per cent of comments on human pull requests name an item outside the change (intervals 3.1 to 3.8, 4.2 to 4.8 and 5.3 to 6.3). For comments on agent pull requests, it predicts 3.1, 4.7 and 6.7 per cent (intervals 2.7 to 3.6, 4.3 to 5.1 and 6.0 to 7.6). Across these deciles, size moves either share by more than the gap between the arms at any one of them. The predictions move a repository's own mean size with its size, so they combine the within-repository and the between-repository terms; the deciles are taken over comments. Because the rise lies mainly between small and mid-sized repositories (Section 4.2), the log-linear prediction at the ninth decile extends it to sizes where the data show it only imprecisely. The instrument's recall is lower on human pull requests, so these predicted levels are not directly comparable between the arms. The predictions also compare repositories, not one repository as it grows.

PR size may lie on the path from authorship to what reviewers write, so the specifications of Table B2 that hold it fixed estimate the association net of that path, not the whole of it.

\setcounter{table}{0}\renewcommand{\thetable}{F\arabic{table}}
\setcounter{figure}{0}\renewcommand{\thefigure}{F\arabic{figure}}
\section*{Appendix F Threats to validity in detail}
\addcontentsline{toc}{section}{Appendix F Threats to validity in detail}
\subsection*{F.1 Construct validity}
\addcontentsline{toc}{subsection}{F.1 Construct validity}
\textbf{What the instrument counts.} The outcome records that a reviewer named an existing item outside the change; it does not record why. A comment counts when it names a file, directory, module or defined symbol that exists at the pull request's base commit, the state the contributor started from, and that lies outside the files the pull request itself had changed up to the commit the comment was written on. A comment that points outside the change without writing a name the repository index can resolve is not counted. Two checks bear on what the named items mean. Named files are later changed in the same pull request more often than other files of their directory, and, in the judges' labels, most comments that name them direct the contributor to the item. The study does not measure whether the contributor acted on the comment, or whether naming an item reflects a need for project knowledge.

\textbf{Precision and recall.} The instrument is usually right when it flags a comment, but it misses many references. Against the validation's reference labels, precision, the share of the instrument's hits that are correct, is 81.6 per cent in the agent arm and 78.8 per cent in the human arm. Recall, the share of true references the instrument finds, is lower: 59.3 per cent in the agent arm and 38.8 per cent in the human arm, mostly against people's labels. The instrument resolves names written as code or as paths, so it misses references written in prose and references to items the index does not define. Its uniqueness limits (a file name carried by at most three paths, a symbol defined in at most five files) also reject more genuine references in larger trees, where names repeat more often. Recall should therefore fall with size, which would bias the measured size gradient toward 1. Precision, however, falls with size in the human arm (below), which adds to that arm's measured gradient, and the error-corrected estimate is lower than the uncorrected one in the same specification (1.22 against 1.56), so the net direction of the measurement error is unknown.

Lower recall in one arm makes that arm's share look smaller, which biases a comparison of levels between the arms toward the other arm. On the odds-ratio scale, it biases a comparison of size gradients only where recall changes with size differently in the two arms. On the percentage-point scale, even a constant difference in recall scales each arm's slope: in an illustration that assumes precision and recall constant over size and carries no interval, rescaled by each arm's precision over recall, the net slopes of Appendix E become 2.80 for agent and 2.63 for human pull requests, against 2.03 and 1.30 measured. The validation holds 21 counted misses across the six cells of arm and size third, too few to tell whether recall changes with size differently in the two arms. Precision falls with size in the human arm (84.2, 78.1 and 76.5 per cent) and is flat in the agent arm (82.5, 83.7 and 80.9 per cent). False positives therefore raise the measured share in larger repositories, from the smallest to the largest third by about 2 per cent of the share in the agent arm and 10 per cent in the human arm. They add little to the agent gradient and more to the human one, which pushes the arm ratio toward 1.

\textbf{A rough precision correction of the arm ratio.} With recall held constant, the measured share is proportional to the true share divided by precision, so a precision that falls with size adds its fall in log precision per tenfold increase to the measured gradient; at shares this small the log odds are close to the log of the share. From the smallest to the largest size third, log precision falls by 0.02 in the agent arm and 0.10 in the human arm. We take the two thirds to lie about 1.5 tenfold steps of size apart, an assumption close to the distance between their typical sizes. Removing the precision gradient from both arms would then multiply the arm ratio by about 1.05: the pre-specified ratio would become 1.21, and the upper end of its 90 per cent interval 1.31, above the bound of 1.25. The calculation treats the precision estimates by third as known, although their intervals are wide, and leaves recall unchanged.

\textbf{Reference labels.} The validation's reference rests on facts from GitHub and on judgements by language models, part of which people checked. The facts come from the GitHub API, a source the instrument does not read: the matched item exists at the base commit, a matched symbol occurs in the file it resolved to, and the item lies outside the change. Whether a comment refers to the matched item, and whether an unflagged comment names one, was judged by models from three families: Gemma (Google) and Claude Sonnet (Anthropic) as judges, blind to arm and size, and Qwen (Alibaba) as adjudicator where the two disagreed. Model judges align with human judgement unevenly from one software engineering task to another (Ahmed et al., 2025; Wang et al., 2025). People made 190 checks of 186 labels and agreed with the final model label on 95 per cent. On the first set of sheets the label was shown, which can anchor a reader. On the second, each label shown was the model's or its reverse at random, so the reader had to judge every item, and its single reader agreed with the model on 92 per cent. The other labels rest on the model reference.

\textbf{Common library names.} Names of widely used libraries produce false matches, and these raise the measured share. When we read a sample of flagged comments, we found this false positive recurring: the comment names a widely used library, and the name resolves to a file, module or symbol of the same name in the repository. Name matching is known to produce false links when a code name is also a common word or a concept (Bacchelli et al., 2010; Dagenais and Robillard, 2012). The validation's precision counts these matches in both arms. The negative controls count them as chance matches to the extent that partner repositories carry the same names.

\textbf{Error correction.} Once the instrument's measured error is corrected, the estimates are too imprecise to confirm or rule out the measured gradients. Design-based supervised learning corrects an estimate from an imperfect measure with a validation sample drawn with known probabilities (Egami et al., 2023). Corrected in this way, the agent-arm size odds ratio, the change in the odds of an outside reference per tenfold increase in size, is 1.22 (0.88 to 1.69), against 1.56 (1.30 to 1.87) uncorrected in the same specification, and the ratio of the agent to the human size odds ratio, which we call the arm ratio below, is 0.80 (0.32 to 2.00). The corrections rest on 300 and 493 labelled comments, so their intervals are wide and neither excludes 1. The corrected models are also simpler than the pre-specified ones: the gradient model has no agent or month effects, and the arm model carries repository means of arm and size only.

\textbf{Arm assignment.} Some human-arm pull requests may have been written with an agent, and these would understate any agent difference. We identify agent pull requests by author account, head-branch prefix and signatures in the description. Agent use that leaves none of these traces is missed (Robbes et al., 2026), and developers do not always declare AI-generated code (Kashif et al., 2026). The human arm is therefore the arm without such markers. Agent-written pull requests in it draw the two arms' gradients together, so the comparison understates any agent difference. Without data on how many there are, we cannot bound the understatement; under a simple mixture on the log-odds scale, a true ratio of 1.25 would appear as 1.15 only if about 37 per cent of human-arm pull requests were agent-written. We say human for short throughout. The agent arm itself mixes two authorship modes: pull requests opened from an agent's bot account, and pull requests opened under the developer's own account, which carry 53 per cent of the RQ1 and RQ2 comments and often have an insider as author (Section 3.2).

\textbf{Reviewers.} Some comments counted as reviewers' may come from automated accounts or from delegators, and we report the estimates with each group left out. A reviewer is a user account whose login does not end in [bot]. Some accounts post automated or AI-generated reviews under other names. Leaving out logins that mark a bot or app gives a size odds ratio of 1.36 and an arm ratio of 1.15. Comments whose text is formatted as AI review tools format their output (severity labels, warning emoji, collapsible blocks, headed sections) are 2,129 agent-arm comments or 2.3 per cent, with an outside-reference rate of 19.6 per cent; leaving them out gives a size odds ratio of 1.32 (interval 1.13 to 1.54). Two kinds remain: AI-generated reviews posted from personal accounts, and review apps that post under ordinary-looking names, which matching by name misses and the collected comments do not mark.

On pull requests opened by an agent's bot account, the delegator comments as a reviewer and writes 36 per cent of agent-arm comments. The delegator's role has no counterpart in the human arm, which is why we also report the estimates without these comments. We identify delegators from assignees, review requests and the description, a rule that also catches some reviewers who were only asked to review. Without these comments the size odds ratio is 1.49, and the arm ratio is 1.17 in the pre-specified model and 1.02 with the agent group and the month held fixed. Of the 33,196 agent-arm comments this rule marks, 23,654 come from delegators found as an assignee or in a ``Requested by'' line, the sources that name who set the agent to work. Leaving out only those, the size odds ratio is 1.45 (interval 1.21 to 1.75) and the arm ratio 1.17 (interval 1.05 to 1.30).

Delegators' comments are not spread evenly over the agents. Under the broad rule, they are 81.8 per cent of the comments on Copilot pull requests, against 44.6 per cent for Devin, 1.0 for Cursor, 18.2 for Google Jules, 1.3 for Claude Code and 0.0 for OpenAI Codex. Under the narrow rule, the shares are 61.5 per cent for Copilot and 17.2 per cent for Devin. This may change what Copilot's reviewers name.

Comments addressed to an agent by its handle are excluded from the primary analyses; they are 17 per cent of agent-arm comments. Restoring them gives a share of 6.0 per cent, a size odds ratio of 1.34 and an arm ratio of 1.15. Addressed comments are most common on Copilot pull requests, at 33.7 per cent of their comments. They are 0.3 per cent for Claude Code, 0.1 for OpenAI Codex, 0.6 for Devin, 0.5 for Cursor and 3.4 for Google Jules, and 0.04 per cent on human pull requests. Their outside-reference rate is 3.9 per cent on Copilot, 5.6 per cent on Claude Code and 7.7 per cent on OpenAI Codex pull requests, the last resting on 13 comments. Devin and human pull requests carry few addressed comments (42 and 159, against 18,442 on Copilot and 108 on Claude Code pull requests); on them the rate is 4.8 and 2.5 per cent. None of the 8 Cursor and 6 Google Jules addressed comments names an outside item.

\textbf{Size and the symbol index.} Repository size and the index of definitions do not match exactly the code a reviewer could name at the base commit. Size counts the code files of the whole tree at the base commit, vendored directories included, while the instrument excludes vendored paths from matching. Repositories that vendor code are therefore recorded as larger than the code a reviewer can name.

Symbol definitions are read at the default branch's head and restricted to files that exist at the base commit. A symbol added to such a file after the base commit can therefore still resolve, and one moved out of it by the head cannot. Definitions are recognised by their declaring keyword. A method declared without one (a Go method with a receiver; a Java, C\#, C++ or Scala method; a Kotlin fun) therefore resolves only through its file's name. Files outside the code extensions, such as Fortran sources, are not read for definitions.

For RQ4, the pre-specified set of files each agent reads is fixed over the window. By GitHub's public changelog, Copilot's coding agent also reads AGENTS.md from August 2025, and counting it gives a ratio of 1.00 (0.73 to 1.36). The same announcement (GitHub's changelog of 28 August 2025) lists CLAUDE.md and GEMINI.md among the files Copilot's coding agent reads; this sensitivity adds AGENTS.md only, so it may still undercount Copilot's context files. We could not date when Cursor began reading AGENTS.md, so its file set stays as pre-specified; Cursor pull requests carry 1,655 of the RQ4 comments, too few for dating its file set to move the RQ4 estimate.

\textbf{Known groups.} The known-groups hypothesis is not supported, so the check gives no support to reading a named item as a transfer of project knowledge. The check compares groups expected to differ: comments by insider reviewers, those with an owner, member or collaborator association, were expected to carry outside references more often. The odds ratio for insider reviewers is 0.89 (0.68 to 1.17) in the agent arm, a failure as the plan defines it, and 1.00 (0.90 to 1.12) in the human arm, above 1 by a negligible margin; neither result supports the hypothesis. Compared within repositories, the odds ratios are 1.04 (0.81 to 1.33) and 1.01 (0.92 to 1.11). Without the delegators, most of them insiders, the agent-arm odds ratio is 1.15 (0.85 to 1.56). The check's premise may also be weak, since reviewers outside the owner, member and collaborator associations include experienced contributors. The association is GitHub's author\_\allowbreak{}association field as read at collection, not at the time of the pull request, and it depends on whether members make their organisation membership public; no comment in the data carries GitHub's first-time-contributor labels, which suggests the field reflects later standing. Both blur insiders and others, so the contrasts built on the field are attenuated toward 1.

\subsection*{F.2 Internal validity}
\addcontentsline{toc}{subsection}{F.2 Internal validity}
\textbf{Agent, month and pull-request size.} We read the arm comparison as an association with authorship, not an effect of it. The pre-specified RQ3 model leaves out the agent and the month, and both vary with repository size. Holding them fixed brings the arm ratio to 0.99 (0.91 to 1.09). Comparing agent and human pull requests within the same repository and month, with PR size and the agent group held fixed, brings it to 1.06 (0.92 to 1.22). Holding PR size fixed alone leaves it at 1.15; agent and human pull requests in the comparison set are of similar size, counted once per pull request (mean log10 lines 2.29 and 2.29; median changed files 5 and 6). From the smallest to the largest third of repositories, the mean is 2.11, 2.24 and 2.31 for agent and 2.13, 2.29 and 2.32 for human pull requests, and the median number of changed files is 3, 5 and 5 for agent and 4, 6 and 6 for human pull requests. PR size may itself lie on the path from authorship to what reviewers write, so holding it fixed estimates the association net of that path.

Holding the month fixed has a cost. If agents changed over the window, it could also remove part of an agent effect, as adjusting for a mediator, a variable on the path from cause to outcome, does (Schisterman et al., 2009). We therefore also report the comparison within each period. Separately, the change could not be fetched at the comment's commit, and fell back to the pull request's head, for 1.2 per cent of agent-arm and 2.0 per cent of human-arm comments.

\textbf{Author standing.} The split of human pull requests by author standing (Appendix B) rests on GitHub's author\_\allowbreak{}association field, read at collection rather than when the pull request was opened and dependent on whether members make their membership public. Both blur insider and non-member authors, so the difference between their gradients is attenuated toward 1. GitHub marks almost all non-member authors in the comparison set as earlier contributors, so the data do not isolate newcomers, who meet social barriers on a first contribution (Steinmacher et al., 2015).

\textbf{Repository properties.} The agent-arm size gradient compares repositories with one another, so properties that grow with size could stand behind it. Across repositories, size correlates with stars at r = 0.41 and with the number of distinct reviewers at r = 0.51. Larger repositories also differ in popularity, age, number of contributors and review practice. Popularity may change how agent pull requests are received (Bhavnagri, 2026), and agent adoption is higher in larger projects (Robbes et al., 2026). With the repository's stars, age and human pull-request authors added, the size odds ratio is 1.43 (1.21 to 1.69), so these three measures do not explain the gradient. Other properties that grow with size, such as review practice, remain. The arm comparison within repositories removes every repository property that acts on both arms alike.

\textbf{Selection on review.} We study only pull requests with at least one review thread, and this selection differs little between the arms. Selecting on review can bias a comparison conditioned on it (Elwert and Winship, 2014). Among all pull requests in the window, 30 per cent of agent and 30 per cent of human pull requests received a review thread. From the smallest to the largest third of repositories, the share rises from 25 to 31 per cent in the agent arm and from 28 to 31 per cent in the human arm. The gradient and the arm comparison therefore describe reviewed pull requests. Coding unreviewed pull requests as having no reference would mix the decision to review with what reviewers write, so we model reviewed pull requests and report the share reviewed instead. Reviewed pull requests carry 3.9, 3.6 and 3.4 eligible comments on average in the agent arm, from the smallest to the largest third of repositories, and 3.6, 3.6 and 3.6 in the human arm. With one row per reviewed pull request, the size odds ratio is 1.27 (1.14 to 1.42), and with the number of comments on the pull request added it is 1.36, so the gradient does not rest on how comments are spread over pull requests. At the pull-request level, whether any comment names an outside item gives an arm ratio of 1.05, because agent pull requests in larger repositories receive fewer comments; counting the outside references with the comments as exposure gives 1.24 (interval 1.09 to 1.40), so the excess per comment persists at that level.

\textbf{Size changes over time.} In the pre-specified RQ3 model, the comment-level size terms describe how a repository grows over time, and they are not comparable with the gradient across repositories. They are identified from how a repository's size changes between base commits, together with everything else that changes over the same months. Per tenfold change, these within-repository size odds ratios are 8.04 for human and 9.26 for agent pull requests; the agent figure is the human within-repository slope times the arm interaction. The repository mean of size carries an odds ratio of 0.18 (interval 0.10 to 0.32; z = -6.0), so the within-repository and between-repository slopes differ, as the Mundlak test of their equality shows. The arm interaction rests mainly on how the gap between agent and human pull requests differs across repositories of different size.

\subsection*{F.3 External validity}
\addcontentsline{toc}{subsection}{F.3 External validity}
\textbf{Frame.} The results describe public open-source repositories on GitHub in which the six agent groups opened pull requests between December 2024 and September 2026. Repositories without agent pull requests, private and industrial repositories, and other agents are outside this scope. The frame is built from the repositories of the AIDev agent arm, whose curated subsets keep only repositories above a star threshold (Li et al., 2025, 2026); the frame's original 1,690 repositories came from the curated subset of repositories with more than 100 stars. For the power of the gradient analysis, it was enlarged with other AIDev repositories of at least 100 stars to 4,088 available repositories (D1, Section 3.1), so the frame is not limited to the curated subset's repositories with agent review comments. The human arm is a seeded sample of up to 100 pull requests per repository and, for the arm comparison, at most 300 comments per repository, so that busy repositories do not dominate.

\textbf{Unit of measurement.} We measure only the inline comment that opens a review thread. Thread replies, review summaries and comments in the pull request's conversation are out of scope, and how much of reviewers' pointing goes through them may differ by arm: Copilot's coding agent takes most requests through comments that mention it; the other agents are addressed this way far less often (Appendix F.1), and Codex and Claude Code sessions are often driven from the developer's own environment. The shares and gradients describe thread-opening inline comments.

Comments were excluded for three reasons. Of 486,378 eligible comments, we excluded 5 duplicate records, 60 comments the instrument could not measure and 2,424 comments whose repository had zero or unknown size at the base commit (197 in the agent arm, 2,227 in the human arm), leaving 483,889. The size models cannot include the sizeless comments. Their outside-reference rate is 1.0 per cent in the agent arm and 0.3 per cent in the human arm, below the measured share; including them gives an RQ1 share of 6.5 per cent.

\textbf{Collection cap.} The pull-request listings of some large repositories were capped, and we completed them before any analysis. These listings stopped at the collector's cap of 20,000 pull requests and covered only the later part of the window. We completed these repositories by date-sliced search over the older part of the window and drew their selection again with the same rule and seed (D2). The completion uses GitHub's search, a different route from the listing used for the other repositories. The completeness audit compares the pull requests collected with GitHub's search count per repository: the shortfall is 0.30 per cent, within the pre-specified threshold of 1 per cent.

\subsection*{F.4 Conclusion validity}
\addcontentsline{toc}{subsection}{F.4 Conclusion validity}
\textbf{Estimator and bound.} The arm comparison's equivalence verdict depends on the bound and on the estimator. The pre-specified model uses the Mundlak device, which adds each repository's means as terms (Mundlak, 1978; Chamberlain, 1980). Its equivalence verdict holds at the bound of 1.25, with a 90 per cent upper bound of 1.2490, and not at narrower bounds: at bounds of 1.20, 1.15 and 1.10, the equivalence test gives p = 0.201, 0.507 and 0.819. These tests describe the precision achieved and are not part of the confirmatory inference. With repository fixed effects in its place, which compare only within each repository, the ratio is 1.17 (1.02 to 1.34) and the equivalence test gives p = 0.1731. The fixed-effects models are fitted by one implementation (pyfixest). A fixed-effects logit keeps only the repositories whose comments vary in the outcome: 1,435 of 1,845 repositories and 331,117 comments for RQ3; 6,819 of 24,865 repository-month cells and 182,607 of 348,721 comments for the comparisons within the same repository and month; 156,105 of 178,398 comments for the fixed-effects model before November 2025; and 728 of 2,034 repositories (77,707 of 92,042 comments) and 1,705 of 2,788 repositories (345,418 of 373,069 comments) for the known-groups checks within repositories. Cells and repositories are counted from the data as those whose outcome varies, and comments as the fitted models report them; the repository-month model and the agent-arm known-groups model keep 39 and 4 more comments than those groups hold. The fixed-effects interaction has a standard error of 0.070 on the log-odds scale, against 0.050 in the Mundlak model. The gap does not come from the repositories the fixed-effects logit drops: fitted on the same 1,435 repositories, the Mundlak model gives 1.16 with a standard error of 0.052. It comes from the estimator. The fixed-effects logit estimates an intercept for each repository and uses only the variation within each, while the Mundlak model replaces those intercepts with a linear function of the repository means, which in a nonlinear model also draws on differences between repositories. The comparisons within the same repository and month are equivalent to 1, except without the delegators' comments, where the verdict is inconclusive (Table B2); with a square of size per arm, they are not, and the ratio falls with size, from 1.45 at 300 code files to 0.98 at 3,000 (Table B3); with a slope per size class, the excess is nominally detected only below 300 code files, only with the delegators' comments kept, and not after Holm adjustment over the three size classes (Table B4). On the repositories the fixed-effects logit keeps, the pre-specified model's equivalence test gives p = 0.0678, so its verdict at the bound also depends on the sample. It also depends on the functional form of size: with a square of size shared by both arms or separate for each, the pre-specified model's equivalence test fails (p = 0.0695 and 0.4221; Table B2). The decisions rest on the two pre-specified tests, Holm-adjusted across the gradient and the arm comparison, with equivalence tests at the smallest effects of interest (Lakens, 2017); the sensitivities carry no decision.

\textbf{Chance matches.} The negative controls remove coincidental matches, not the matchable names that a larger repository's own code adds around a change. Each control looks a comment's names up in partner repositories of another owner: the next repository in size, and five random partners of the same language and similar size. A partner is read at its head with nothing excluded as changed, and partners are paired by the number of paths at the head, not by code files at the base commit. The net contrasts are excess over chance under these assumptions and one more: that chance matching adds the same amount to the measured outcome as to the control. The control is a negative-control outcome in the sense of Lipsitch et al. (2010), who use such outcomes to detect bias rather than to subtract it.

We also re-paired the next-partner control. Re-paired on code files at the head, the measure our size counts at the base (D20), the next-partner control gives a net agent-arm gradient of 1.71 percentage points per tenfold increase (interval 0.93 to 2.49) and a net arm difference of 0.81 (0.33 to 1.29), against 1.89 and 0.91 paired on all paths.

Some repositories could not be matched closely in size. For a repository with fewer than five eligible partners within 15 per cent of size, mostly repositories in rare languages and the largest repositories of their language, the band widens to 30, 50 and then 100 per cent, and the width each comment was matched at is recorded (D18). Restricted to repositories matched within 15 per cent, the net agent-arm gradient is 2.88 percentage points per tenfold size (1.56 to 4.19), against 2.00 (1.08 to 2.93) with the widened bands. The net arm difference is 1.29 percentage points (0.55 to 2.02), against 0.90 (0.40 to 1.40).

\textbf{Design diagnosis.} Power, type I error, coverage and bias were simulated (200 simulations per scenario) for the earlier round's frame and estimators, not for this frame, its estimators or its selection. The equivalence test at narrower bounds (above) shows the precision achieved instead.

\textbf{Departures from the pre-specified plan.} Table A1 (Appendix A) lists every departure from the plan and when we decided it; the verdicts reported as decisions are those of the pre-specified models. D19 to D22 were decided after this run's estimates had been seen and add sensitivities and checks, except D21, which restores the pre-specified definition of the change; the pre-specified verdicts are the same before and after it, and a table of the pre-specified estimates before and after D21 is in the replication package. The pre-specified RQ3 model carries the size main effect that the frozen formula omitted (D9). The frozen formula, the earlier round's, leaves out the size main effect while size varies within repositories. Its interaction therefore absorbs part of the size gradient itself. It gives 1.43, and its 90 per cent interval, 1.2509 to 1.6319, lies wholly above the bound (Table B2). The two formulas therefore reach different verdicts, and both are reported.

\section*{References}
\addcontentsline{toc}{section}{References}
\begin{list}{}{\setlength{\leftmargin}{1.6em}\setlength{\itemindent}{-1.6em}\setlength{\itemsep}{2pt}\small\raggedright}
\item Ahmed T, Devanbu P, Treude C, Pradel M (2025) Can LLMs Replace Manual Annotation of Software Engineering Artifacts? In: 2025 IEEE/\allowbreak{}ACM 22nd International Conference on Mining Software Repositories (MSR), pp 526–538. \url{https://doi.org/10.1109/msr66628.2025.00086}
\item Aickin M, Gensler H (1996) Adjusting for multiple testing when reporting research results: the Bonferroni vs Holm methods. American Journal of Public Health 86(5):726–728. \url{https://doi.org/10.2105/ajph.86.5.726}
\item Ali N, Baker S, O’Crowley R, Herold S, Buckley J (2018) Architecture consistency: State of the practice, challenges and requirements. Empirical Software Engineering 23(1):224–258. \url{https://doi.org/10.1007/s10664-017-9515-3}
\item Allamanis M, Barr ET, Bird C, Sutton C (2014) Learning natural coding conventions. In: Proceedings of the 22nd ACM SIGSOFT International Symposium on Foundations of Software Engineering, pp 281–293. \url{https://doi.org/10.1145/2635868.2635883}
\item Arabat A, Sayagh M (2026) Toward Instructions-as-Code: Understanding the Impact of Instruction Files on Agentic Pull Requests. In: Proceedings of the 23rd International Conference on Mining Software Repositories, pp 969–973. \url{https://doi.org/10.1145/3793302.3793601}
\item Bacchelli A, Bird C (2013) Expectations, outcomes, and challenges of modern code review. In: 2013 35th International Conference on Software Engineering (ICSE), pp 712–721. \url{https://doi.org/10.1109/icse.2013.6606617}
\item Bacchelli A, Lanza M, Robbes R (2010) Linking e-mails and source code artifacts. In: Proceedings of the 32nd ACM/\allowbreak{}IEEE International Conference on Software Engineering - Volume 1, pp 375–384. \url{https://doi.org/10.1145/1806799.1806855}
\item Bairi R, Sonwane A, Kanade A, C VD, Iyer A, Parthasarathy S, Rajamani S, Ashok B, Shet S (2024) CodePlan: Repository-Level Coding using LLMs and Planning. Proceedings of the ACM on Software Engineering 1(FSE):675–698. \url{https://doi.org/10.1145/3643757}
\item Baltes S, Cheong M, Treude C (2026) ``An Endless Stream of AI Slop'': How Developers Discuss the Burden of AI-Assisted Software Development. IEEE Software. Online first. \url{https://doi.org/10.1109/ms.2026.3730626}
\item Beller M, Bacchelli A, Zaidman A, Juergens E (2014) Modern code reviews in open-source projects: which problems do they fix? In: Proceedings of the 11th Working Conference on Mining Software Repositories, pp 202–211. \url{https://doi.org/10.1145/2597073.2597082}
\item Bhavnagri P (2026) AI Coding Agents on GitHub: Repository Popularity as a Hidden Confounder in the Acceptance of 2.4 Million Agent-Authored Pull Requests. International Journal for Research in Applied Science and Engineering Technology 14(9):165–176. \url{https://doi.org/10.22214/ijraset.2026.84804}
\item Bosu A, Greiler M, Bird C (2015) Characteristics of Useful Code Reviews: An Empirical Study at Microsoft. In: 2015 IEEE/\allowbreak{}ACM 12th Working Conference on Mining Software Repositories, pp 146–156. \url{https://doi.org/10.1109/msr.2015.21}
\item Brahmi Z, Ouni A, Sayagh M, Saied MA (2026) Characterizing Self-Admitted Technical Debt Generated by AI Coding Agents. In: Proceedings of the 23rd International Conference on Mining Software Repositories, pp 852–856. \url{https://doi.org/10.1145/3793302.3793588}
\item Calderon N, Reichart R, Dror R (2025) The Alternative Annotator Test for LLM-as-a-Judge: How to Statistically Justify Replacing Human Annotators with LLMs. In: Proceedings of the 63rd Annual Meeting of the Association for Computational Linguistics (Volume 1: Long Papers), pp 16051–16081. \url{https://doi.org/10.18653/v1/2025.acl-long.782}
\item Cameron AC, Gelbach JB, Miller DL (2008) Bootstrap-Based Improvements for Inference with Clustered Errors. Review of Economics and Statistics 90(3):414–427. \url{https://doi.org/10.1162/rest.90.3.414}
\item Campbell DT, Fiske DW (1959) Convergent and discriminant validation by the multitrait-multimethod matrix. Psychological Bulletin 56(2):81–105. \url{https://doi.org/10.1037/h0046016}
\item Chamberlain G (1980) Analysis of Covariance with Qualitative Data. The Review of Economic Studies 47(1):225–238. \url{https://doi.org/10.2307/2297110}
\item Chatlatanagulchai W, Li H, Kashiwa Y, Reid B, Thonglek K, Leelaprute P, Rungsawang A, Manaskasemsak B, Adams B, Hassan AE, Iida H (2026) Agent READMEs: An Empirical Study of Context Files for Agentic Coding. ACM Transactions on Software Engineering and Methodology. Online first. \url{https://doi.org/10.1145/3840295}
\item Cohen J (1960) A Coefficient of Agreement for Nominal Scales. Educational and Psychological Measurement 20(1):37–46. \url{https://doi.org/10.1177/001316446002000104}
\item Cronbach LJ, Meehl PE (1955) Construct validity in psychological tests. Psychological Bulletin 52(4):281–302. \url{https://doi.org/10.1037/h0040957}
\item Dagenais B, Robillard MP (2012) Recovering traceability links between an API and its learning resources. In: 2012 34th International Conference on Software Engineering (ICSE), pp 47–57. \url{https://doi.org/10.1109/icse.2012.6227207}
\item Davila N, Nunes I (2021) A systematic literature review and taxonomy of modern code review. Journal of Systems and Software 177:110951. \url{https://doi.org/10.1016/j.jss.2021.110951}
\item Destefanis G, de Souza Santos R, Ortu M, Wessel M (2026) The Dose Makes the Agent: Therapeutic Index Analysis of AI Coding Contributions. In: Proceedings of the 23rd International Conference on Mining Software Repositories, pp 914–918. \url{https://doi.org/10.1145/3793302.3793600}
\item Egami N, Hinck M, Stewart BM, Wei H (2023) Using Imperfect Surrogates for Downstream Inference: Design-based Supervised Learning for Social Science Applications of Large Language Models. In: Advances in Neural Information Processing Systems 36, pp 68589–68601. \url{https://doi.org/10.52202/075280-3000}
\item Ehsani R, Pathak S, Rawal S, Al Mujahid A, Imran MM, Chatterjee P (2026) Where Do AI Coding Agents Fail? An Empirical Study of Failed Agentic Pull Requests in GitHub. In: Proceedings of the 23rd International Conference on Mining Software Repositories, pp 807–811. \url{https://doi.org/10.1145/3793302.3793579}
\item Elwert F, Winship C (2014) Endogenous Selection Bias: The Problem of Conditioning on a Collider Variable. Annual Review of Sociology 40(1):31–53. \url{https://doi.org/10.1146/annurev-soc-071913-043455}
\item Farrag SE (2026) The Productivity-Reliability Paradox: Specification-Driven Governance for AI-Augmented Software Development. arXiv preprint arXiv:2605.01160. \url{https://doi.org/10.48550/arxiv.2605.01160}
\item Fleiss JL (1971) Measuring nominal scale agreement among many raters. Psychological Bulletin 76(5):378–382. \url{https://doi.org/10.1037/h0031619}
\item Gloaguen T, Mündler-Sasahara N, Müller MN, Raychev V, Vechev M (2026) Evaluating AGENTS.md: Are Repository-Level Context Files Helpful for Coding Agents? arXiv preprint arXiv:2602.11988. \url{https://doi.org/10.48550/arxiv.2602.11988}
\item Haider MA, Zimmermann T (2026) Understanding Dominant Themes in Reviewing Agentic AI-authored Code. In: Proceedings of the 23rd International Conference on Mining Software Repositories, pp 742–746. \url{https://doi.org/10.1145/3793302.3793566}
\item Hayes AF, Krippendorff K (2007) Answering the Call for a Standard Reliability Measure for Coding Data. Communication Methods and Measures 1(1):77–89. \url{https://doi.org/10.1080/19312450709336664}
\item Hindi M, Mahmood Y, Mohammed L, Bouktif S, Mediani M (2026) Coding Agents in the Wild: Failure Modes and Rejection Patterns of AI-Generated Pull Requests. IEEE Access 14:83075–83094. \url{https://doi.org/10.1109/access.2026.3696573}
\item Hirao T, McIntosh S, Ihara A, Matsumoto K (2019) The review linkage graph for code review analytics: a recovery approach and empirical study. In: Proceedings of the 2019 27th ACM Joint Meeting on European Software Engineering Conference and Symposium on the Foundations of Software Engineering, pp 578–589. \url{https://doi.org/10.1145/3338906.3338949}
\item Horvitz DG, Thompson DJ (1952) A Generalization of Sampling Without Replacement from a Finite Universe. Journal of the American Statistical Association 47(260):663–685. \url{https://doi.org/10.1080/01621459.1952.10483446}
\item Huang H, Jaisri P, Shimizu S, Chen L, Nakashima S, Rodríguez-Pérez G (2026) More Code, Less Reuse: Investigation on Code Quality and Reviewer Sentiment towards AI-generated Pull Requests. In: Proceedings of the 23rd International Conference on Mining Software Repositories, pp 1024–1028. \url{https://doi.org/10.1145/3793302.3793622}
\item Jimenez CE, Yang J, Wettig A, Yao S, Pei K, Press O, Narasimhan K (2024) SWE-bench: Can Language Models Resolve Real-World GitHub Issues? In: International Conference on Learning Representations (ICLR 2024). arXiv:2310.06770. \url{https://doi.org/10.48550/arxiv.2310.06770}
\item Kashif SM, Liang P, Tahir A (2026) On Developers’ Self-Declaration of AI-Generated Code: An Analysis of Practices. ACM Transactions on Software Engineering and Methodology 35(7):1–37. \url{https://doi.org/10.1145/3771937}
\item Khelifi S, Ouni A, Khemaja M (2026) Behind Agentic Pull Requests: An Empirical Study on Developer Interventions in AI Agent-Authored Pull Requests. In: Proceedings of the 23rd International Conference on Mining Software Repositories, pp 842–846. \url{https://doi.org/10.1145/3793302.3793586}
\item Kononenko O, Baysal O, Guerrouj L, Cao Y, Godfrey MW (2015) Investigating code review quality: Do people and participation matter? In: 2015 IEEE International Conference on Software Maintenance and Evolution (ICSME), pp 111–120. \url{https://doi.org/10.1109/icsm.2015.7332457}
\item Konrad PM, Adam TL, Terrenzi R, Ayvaz S (2026) Architecture Without Architects: How AI Coding Agents Shape Software Architecture. arXiv preprint arXiv:2604.04990. \url{https://doi.org/10.48550/arxiv.2604.04990}
\item Lakens D (2017) Equivalence Tests: A Practical Primer for t Tests, Correlations, and Meta-Analyses. Social Psychological and Personality Science 8(4):355–362. \url{https://doi.org/10.1177/1948550617697177}
\item Larsen OA, Moghaddam MT (2026) Mining Architectural Quality Under Agentic AI Adoption: A Causal Study of Java Repositories. arXiv preprint arXiv:2606.13298. \url{https://doi.org/10.48550/arxiv.2606.13298}
\item Li H, Zhang H, Hassan AE (2025) The Rise of AI Teammates in Software Engineering (SE) 3.0: How Autonomous Coding Agents Are Reshaping Software Engineering. arXiv preprint arXiv:2507.15003. \url{https://doi.org/10.48550/arxiv.2507.15003}
\item Li H, Zhang H, Hassan AE (2026) AIDev: Studying AI Coding Agents on GitHub. In: Proceedings of the 23rd International Conference on Mining Software Repositories, pp 1029–1033. \url{https://doi.org/10.1145/3793302.3797249}
\item Li R, Soliman M, Liang P, Avgeriou P (2022) Symptoms of Architecture Erosion in Code Reviews: A Study of Two OpenStack Projects. In: 2022 IEEE 19th International Conference on Software Architecture (ICSA), pp 24–35. \url{https://doi.org/10.1109/icsa53651.2022.00011}
\item Lipsitch M, Tchetgen Tchetgen E, Cohen T (2010) Negative Controls: A Tool for Detecting Confounding and Bias in Observational Studies. Epidemiology 21(3):383–388. \url{https://doi.org/10.1097/ede.0b013e3181d61eeb}
\item Mäntylä MV, Lassenius C (2009) What Types of Defects Are Really Discovered in Code Reviews? IEEE Transactions on Software Engineering 35(3):430–448. \url{https://doi.org/10.1109/tse.2008.71}
\item Martini A, Bosch J (2015) The Danger of Architectural Technical Debt: Contagious Debt and Vicious Circles. In: 2015 12th Working IEEE/\allowbreak{}IFIP Conference on Software Architecture, pp 1–10. \url{https://doi.org/10.1109/wicsa.2015.31}
\item McIntosh S, Kamei Y, Adams B, Hassan AE (2014) The impact of code review coverage and code review participation on software quality: a case study of the qt, VTK, and ITK projects. In: Proceedings of the 11th Working Conference on Mining Software Repositories, pp 192–201. \url{https://doi.org/10.1145/2597073.2597076}
\item Mohsenimofidi S, Galster M, Treude C, Baltes S (2026) Context Engineering for AI Agents in Open-Source Software. In: Proceedings of the 23rd International Conference on Mining Software Repositories, pp 194–198. \url{https://doi.org/10.1145/3793302.3793350}
\item Mundlak Y (1978) On the Pooling of Time Series and Cross Section Data. Econometrica 46(1):69–85. \url{https://doi.org/10.2307/1913646}
\item Nachuma C, Zibran M (2026) When AI Teammates Meet Code Review: Collaboration Signals Shaping the Integration of Agent-Authored Pull Requests. In: Proceedings of the 23rd International Conference on Mining Software Repositories, pp 717–721. \url{https://doi.org/10.1145/3793302.3793561}
\item Ogenrwot D, Businge J (2026) PatchTrack: A comprehensive analysis of ChatGPT’s influence on pull request outcomes. Empirical Software Engineering 31(5):136. \url{https://doi.org/10.1007/s10664-026-10869-5}
\item Olewicki D, Da Silva L, Sghaier OB, Mujahid S, Amini A, Mah B, Castelluccio M, Habchi S, Khomh F, Adams B (2026) Impact of an LLM-based Review Assistant in Practice: A Mixed Open-/\allowbreak{}Closed-source Case Study. IEEE Transactions on Software Engineering. Online first. \url{https://doi.org/10.1109/tse.2026.3663093}
\item Ottenhof L, Penner D, Hindle A, Lutellier T (2026) How do Agents Refactor: An Empirical Study. In: Proceedings of the 23rd International Conference on Mining Software Repositories, pp 797–801. \url{https://doi.org/10.1145/3793302.3793577}
\item Paixao M, Krinke J, Han D, Ragkhitwetsagul C, Harman M (2021) The Impact of Code Review on Architectural Changes. IEEE Transactions on Software Engineering 47(5):1041–1059. \url{https://doi.org/10.1109/tse.2019.2912113}
\item Pascarella L, Spadini D, Palomba F, Bruntink M, Bacchelli A (2018) Information Needs in Contemporary Code Review. Proceedings of the ACM on Human-Computer Interaction 2(CSCW):1–27. \url{https://doi.org/10.1145/3274404}
\item Passos L, Terra R, Valente MT, Diniz R, das Chagas Mendonça N (2010) Static Architecture-Conformance Checking: An Illustrative Overview. IEEE Software 27(5):82–89. \url{https://doi.org/10.1109/ms.2009.117}
\item Rigby PC, Robillard MP (2013) Discovering essential code elements in informal documentation. In: 2013 35th International Conference on Software Engineering (ICSE), pp 832–841. \url{https://doi.org/10.1109/icse.2013.6606629}
\item Robbes R, Matricon T, Degueule T, Hora A, Zacchiroli S (2026) Agentic Much? Adoption of Coding Agents on GitHub. ACM Transactions on Software Engineering and Methodology. Online first. \url{https://doi.org/10.1145/3822180}
\item Russo D (2026) Govern the Repository, Not the Agent: Measuring Ecosystem-Level Risk in AI-Native Software. arXiv preprint arXiv:2606.28235. \url{https://doi.org/10.48550/arxiv.2606.28235}
\item Schisterman EF, Cole SR, Platt RW (2009) Overadjustment Bias and Unnecessary Adjustment in Epidemiologic Studies. Epidemiology 20(4):488–495. \url{https://doi.org/10.1097/ede.0b013e3181a819a1}
\item Seabold S, Perktold J (2010) Statsmodels: Econometric and Statistical Modeling with Python. In: Proceedings of the 9th Python in Science Conference, pp 92–96. \url{https://doi.org/10.25080/majora-92bf1922-011}
\item Song F, Agarwal A, Wen W (2026) The Impact of Generative AI on Collaborative Open-Source Software Development: Evidence from GitHub Copilot. Information Systems Research. Online first. \url{https://doi.org/10.1287/isre.2024.1361}
\item Steinmacher I, Conte T, Gerosa MA, Redmiles D (2015) Social Barriers Faced by Newcomers Placing Their First Contribution in Open Source Software Projects. In: Proceedings of the 18th ACM Conference on Computer Supported Cooperative Work \& Social Computing, pp 1379–1392. \url{https://doi.org/10.1145/2675133.2675215}
\item Strauss MC, Schulze S (2026) Why and When Agentic Pull Requests are (not) Accepted: An Exploratory Study. In: Proceedings of the 23rd International Conference on Mining Software Repositories, pp 767–771. \url{https://doi.org/10.1145/3793302.3793571}
\item Sun K, Kuang H, Baltes S, Zhou X, Zhang H, Ma X, Rong G, Shao D, Treude C (2026) Does AI Code Review Lead to Code Changes? A Case Study of GitHub Actions. IEEE Transactions on Software Engineering 52(7):2111–2126. \url{https://doi.org/10.1109/tse.2026.3688237}
\item Thongtanunam P, McIntosh S, Hassan AE, Iida H (2016) Revisiting code ownership and its relationship with software quality in the scope of modern code review. In: Proceedings of the 38th International Conference on Software Engineering, pp 1039–1050. \url{https://doi.org/10.1145/2884781.2884852}
\item Wang D, Xiao T, Thongtanunam P, Kula RG, Matsumoto K (2021) Understanding shared links and their intentions to meet information needs in modern code review: A case study of the OpenStack and Qt projects. Empirical Software Engineering 26(5):96. \url{https://doi.org/10.1007/s10664-021-09997-x}
\item Wang R, Guo J, Gao C, Fan G, Chong CY, Xia X (2025) Can LLMs Replace Human Evaluators? An Empirical Study of LLM-as-a-Judge in Software Engineering. Proceedings of the ACM on Software Engineering 2(ISSTA):1955–1977. \url{https://doi.org/10.1145/3728963}
\item Watanabe M, Li H, Kashiwa Y, Reid B, Iida H, Hassan AE (2026) On the Use of Agentic Coding: An Empirical Study of Pull Requests on GitHub. ACM Transactions on Software Engineering and Methodology 35(10):1–24. \url{https://doi.org/10.1145/3798166}
\item Wooldridge JM (2019) Correlated random effects models with unbalanced panels. Journal of Econometrics 211(1):137–150. \url{https://doi.org/10.1016/j.jeconom.2018.12.010}
\item Xia CS, Deng Y, Dunn S, Zhang L (2025) Demystifying LLM-Based Software Engineering Agents. Proceedings of the ACM on Software Engineering 2(FSE):801–824. \url{https://doi.org/10.1145/3715754}
\item Yang J, Jimenez C, Wettig A, Lieret K, Yao S, Narasimhan K, Press O (2024) SWE-agent: Agent-Computer Interfaces Enable Automated Software Engineering. In: Advances in Neural Information Processing Systems 37, pp 50528–50652. \url{https://doi.org/10.52202/079017-1601}
\item Zhang F, Chen B, Zhang Y, Keung J, Liu J, Zan D, Mao Y, Lou JG, Chen W (2023) RepoCoder: Repository-Level Code Completion Through Iterative Retrieval and Generation. In: Proceedings of the 2023 Conference on Empirical Methods in Natural Language Processing, pp 2471–2484. \url{https://doi.org/10.18653/v1/2023.emnlp-main.151}
\end{list}
\end{document}